\documentclass[preprint,12pt]{elsarticle}

\usepackage{amssymb}
\usepackage{graphicx}
\usepackage{caption}
\usepackage{subcaption}

\journal{Nuclear Instruments and Methods in Physics Research Section A}

\begin{document}

\begin{frontmatter}



\title{The Experimental Nuclear Reaction Data (EXFOR): Extended Computer Database  and Web Retrieval System}

\author[label1]{V. V.Zerkin}
\cortext[cor1]{Corresponding author}
\address[label1]{Nuclear Data Section, International Atomic Energy Agency, \\ Vienna International Centre, P.O. Box 100, A-1400 Vienna, Austria}
\author[label2]{B. Pritychenko\corref{cor1}}
\ead{pritychenko@bnl.gov}
\address[label2]{National Nuclear Data Center, Brookhaven National Laboratory, \\ Upton, NY 11973-5000, USA}
\address{}

\begin{abstract}
The EXchange FORmat (EXFOR)  experimental nuclear reaction database and the associated Web interface provide access to the wealth of low- and intermediate-energy nuclear reaction physics data. This resource is based on  numerical data sets and bibliographical information of  $\sim$22,000 experiments since the beginning of nuclear science. 
 The principles of the  Web and database applications development are described.  New  capabilities for the data sets uploads, renormalization, covariance matrix, and inverse reaction calculations  are presented. 

\noindent    The EXFOR database, updated monthly, provides an essential support for nuclear data evaluation, application development, and research activities. 
It is publicly available  at the websites of the   International Atomic Energy Agency Nuclear Data Section, {\it http://www-nds.iaea.org/exfor},  
 the U.S. National Nuclear Data Center, {\it http://www.nndc.bnl.gov/exfor}, and the mirror sites in China, India and Russian Federation. 
\end{abstract}

\end{frontmatter}


\section{Introduction}
\label{sec:Intro}
Since the dawn of the nuclear age, access to the most recent nuclear data  played a crucial role in nuclear physics research and application development. 
These data were originally disseminated as paper copies for designated users and eventually evolved into the present-day computer files distributed over 
the World Wide Web.  The National Nuclear Data Center (NNDC),  Brookhaven National Laboratory (BNL) and Nuclear Data Section (NDS), 
International Atomic Energy Agency (IAEA)   have always been pioneers in nuclear data collection, evaluation, and dissemination. 
They have been providing remote electronic access to their databases and other information since 1986. In this work the authors will concentrate on the present state of storage, 
analysis and worldwide distribution of low- and intermediate-energy nuclear reaction data.

EXFOR experimental nuclear reaction database  \cite{Holden05,Otuka14} is the  world's prime repository for  reaction data compilations. 
The EXFOR  library was established in 1967 at a meeting of the four major nuclear data centers: 
Brookhaven National Laboratory, Upton, NY (Area \# 1, USA, Canada); Nuclear Energy Agency (NEA) Databank, Paris, France (Area \# 2, Western Europe and Japan);  
International Atomic Energy Agency, Vienna, Austria (Area \# 3, Asia, Eastern Europe, Latin America, Africa, Australia and Oceania); and 
Institute of Physics and Power Engineering (IPPE), Obninsk, USSR (Area \#4, Soviet Union) \cite{Holden05}. This project was based on  previously-existing efforts, 
such as the NNDC predecessor's Sigma Center Information Storage and Retrieval System (SCISRS) and the associated format. Currently, the EXFOR project is an international 
collaboration under the auspices of the International Atomic Energy Agency, and 13 nuclear data centers from the USA, 
OECD/Data Bank (France), IAEA/NDS (Austria), Russian Federation, China, Hungary, India, Japan, Kazakhstan, Korea, and Ukraine regularly contribute to it.

\section{EXFOR Database}
\label{sec:X4}

\begin{figure}[]
 \centering
 \includegraphics[width=0.7\textwidth]{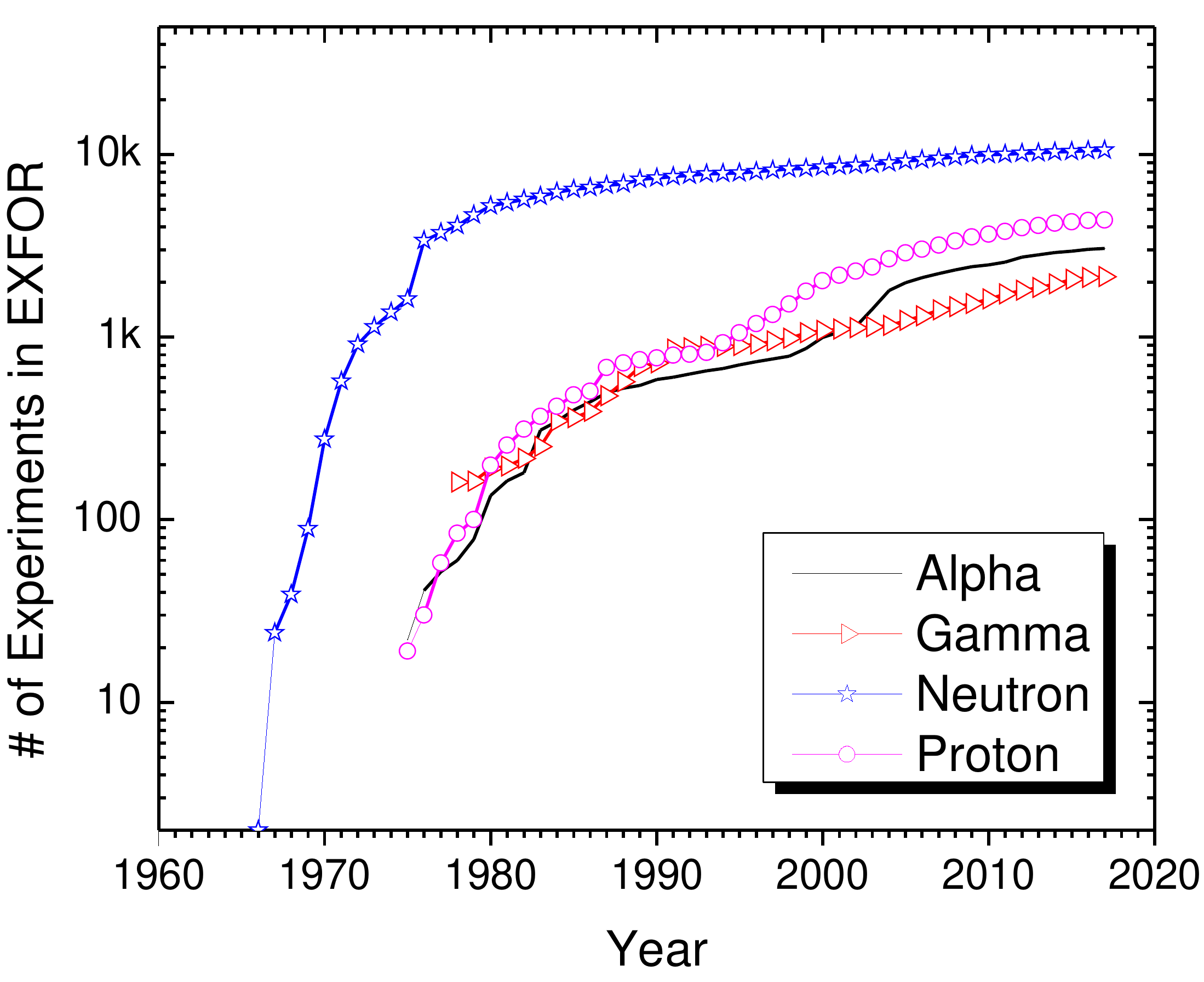}
\caption{Content evolution of the EXFOR database. Neutron reactions have been compiled systematically since the initiation of the EXFOR project, while charged particle and photon (gamma) reactions have 
been covered less extensively. Complementary details are available in Ref. \cite{Pritychenko15}.}
\label{fig:x4}
\end{figure}

The exchange format library was created in support of nuclear energy, fundamental research, and application development activities. 
The initial database scope was limited to neutron-induced reaction quantities. Later, the scope was extended to charged-particle and photon-induced reactions with energies below the pion production threshold. Presently, neutron-, proton-, alpha-, and photon-induced reactions 
constitute 47.9, 19.8, 7.36 and 6.2 $\%$ of database contents, respectively. The historic evolution of EXFOR contents is shown in Fig. \ref{fig:x4}.  
In addition to cross section data, the EXFOR library includes information on incoming particle spectra such as $^{252}$Cf spontaneous fission and $\beta$-delayed neutrons.  
Heavy ion-, electron- and pion-induced reactions with energies up to 1 GeV are compiled by the responsible centers on a voluntary basis. 

The present database is the largest low- and intermediate-energy nuclear reaction library; as of November 27, 2017, it includes 21,885 experiments, 
170,088 data sets, and 15,213,129 data points. The database contains information on 968 targets, 386 incident projectiles, and 2,606 nuclear reactions. 
To further illustrate the volume and diversity of the database and associated computer programming challenges 
we will provide a list of compiled nuclear physics quantities in  Table \ref{Table1}. The table data show that  cross section data sets represent approximately  half of 
the database contents. It also includes substantial contributions from  differential data with respect to angle, partial cross sections, 
resonance parameters, polarization data, fission product yields, and double differential cross section measurements. The total number 
of Table \ref{Table1} compilations exceeds the overall  number of experiments due to a complex structure of compiled entries: 
a single entry often contains multiple data subentries, reaction strings, and reaction combinations.
\begin{table}[htp] 
\begin{center}
\caption[List of EXFOR database compiled nuclear physics quantities.]{List of EXFOR database compiled nuclear physics quantities.}
\vspace{0.2cm}
\label{Table1}
\begin{tabular}{l|l|l|l}
\hline
\# & EXFOR Code & Description & Compilations \\ 
\hline \hline
1 &	CS  &	Cross section data  &	11,204 	\\
2 &	DAP &	Partial differential data with respect to angle &	4,248 	\\
3 &	DA  &	Differential data with respect to angle &	4,234 	\\
4 &	RP  &	Resonance parameters &	1,961 	\\
5 &	CSP &	Partial cross section data &	1,874 	\\
6 &	POL &	Polarization data &	1,108 	\\
7 &	FY  &	Fission product yields &	1,094 \\
8 &	DAE &	Differential data with respect to angle and energy &	1,064 \\
9 &	MFQ &	Fission neutron quantities &	509 \\
10 &	SP  &	Gamma spectra &	461 \\
11 &	RI  &	Resonance integrals &	451 \\
12 &	DE  &	Differential data with respect to energy &	372 \\
13 &	TT  &	Thick target yields &	359 \\
14 &	E   &	Kinetic energies  &	333 \\
15 &	L   &	Scattering amplitudes &	196 \\
16 &	INT &	Cross section integral over incident energy &	185 \\
17 &	PY  &	Product yields &	166 	\\
18 &	NQ  &	Nuclear quantities &	137 	\\
19 &	MLT &	Outgoing particle multiplicities  &	102 	\\
20 &	RR  &	Reaction rates &	98 	\\
21 &	TTD &	Differential thick target yields &	51 \\
22 &	CST &	Temperature dependent cross section data &	30 \\
23 &	SQ  &	Special quantities  &	16 \\
24 &	DEP &	Partial differential data with respect to energy &	12 \\
25 &	TTP &	Partial thick target yields &	7 \\
26 &	COR &	Secondary particle correlations 	& 5 \\
\hline
\end{tabular}
\end{center}
\end{table} 

\begin{figure}[]
 \centering
 \includegraphics[width=0.7\textwidth]{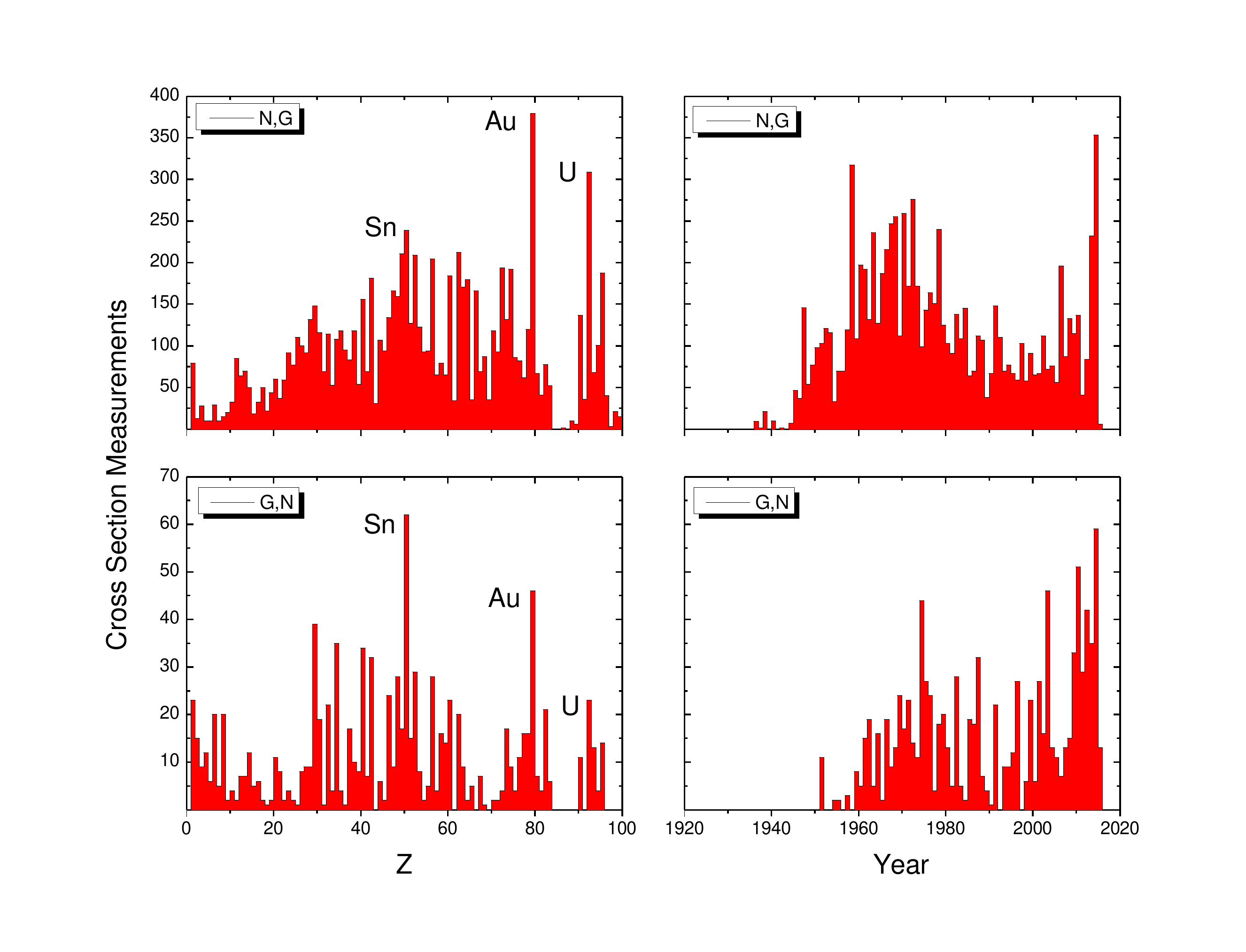}
\caption{EXFOR compilations of nuclear astrophysics relevant neutron capture (n,$\gamma$) and photo-neutron ($\gamma$,n) production reactions vs. target material (z),  
and historical distribution of experimental results are shown in left and right panels, respectively.}
\label{fig:neutron}
\end{figure}

Figure \ref{fig:x4} data highlight the fact that neutron-induced reactions represent a backbone of EXFOR, and these data are essential for many applications. 
The neutron reactions constitute a perfect case of mutually-beneficial relations between the  fundamental and applied sciences. 
For demonstration purposes, the authors will  analyze neutron capture and inverse photoneutron reactions and  the historical evolution of experimental nuclear reaction activities.   
These  direct and inverse reactions  play an essential role in computer modeling of astrophysical {\it s}, {\it r}, and {\it p} processes,  and they are crucial in nuclear reactor calculations. 
Data shown  in the left panels of Fig. \ref{fig:neutron} demonstrate  
that the most frequently-used neutron capture targets are gold (Au), uranium (U), and tin (Sn), 
while neutron-rich tin, gold and uranium are materials of choice for photo-neutron reaction studies. Subsequent examination of the right panels provides 
indication of two trends: large numbers of nuclear reaction measurements in the 1970's \cite{Pritychenko15} 
and a recent increase in number of (n,$\gamma$) and ($\gamma$,n) reactions. Further analysis of the last finding reveals  a substantial contribution of   
Maxwellian-averaged cross section measurements in recent years. 
These experiments tackle nuclear astrophysics problems and provide complementary benchmarks for nuclear reactor modeling.

\subsection{EXFOR Data Compilation}

In the early 1950's,  data compilations have been pioneered 
at the Brookhaven National Laboratory in support of nuclear science and reactor research activities \cite{Holden05,Pearl70}.  
Since 1964 Brookhaven compilations have been stored in the SCISRS  computer system that predated the EXFOR database. 
These experimental data compilation efforts  have always had an important international component. The IAEA Nuclear Data Section has been involved in this work  
 since its creation in 1964. Other early contributors include NEA Data Bank, Paris, France and the Institute of Physics and Power Engineering, 
Obninsk, USSR which were founded in 1964 and 1963, respectively \cite{Holden05}.  In 1969 an agreement on an exchange format was reached between 
four centers and July 1970 was chosen as the starting date for 
transmission of  neutron data between the partners in the EXFOR format.  The effort to translate all of
the existing experimental  nuclear reaction data sets, which were coded in the SCISRS system format, 
 would be done later \cite{Holden05}.  

In recent years, the EXFOR compilations became even more popular worldwide, and a large number of institutions have joined. In fact, EXFOR represents 
one of the oldest continuously-operated scientific collaborations. Since its creation, the EXFOR project has relied heavily on computer technologies 
available at the time. Historically, the data compilations were performed using pencil and paper, and they have gradually developed into the present day computer 
operation that includes modern relational database and Web servers, EXFOR compilation editors, plot digitizers, and optical character recognition technologies. 
Since the beginning, reaction data  compilations  were  performed on a timely basis, and archived in close cooperation with the researchers. 
Many scientists have contributed  their original data sets to the EXFOR library. 
Unfortunately, this mode of operation has not always worked perfectly, and several important data sets were missed 
and the NRDC network members are working on their recovery.  In addition, the NRDC community has addressed the issues with the problematic and incorrectly entered data compilations. 
The database contents were reanalyzed and corrected by the members of the Subgroup 30 of the Working Party on International Evaluation Co-operation of the NEA Nuclear Science Committee \cite{SG30}.

\begin{figure}[]
 \centering
 \fbox{\includegraphics[width=0.95\textwidth]{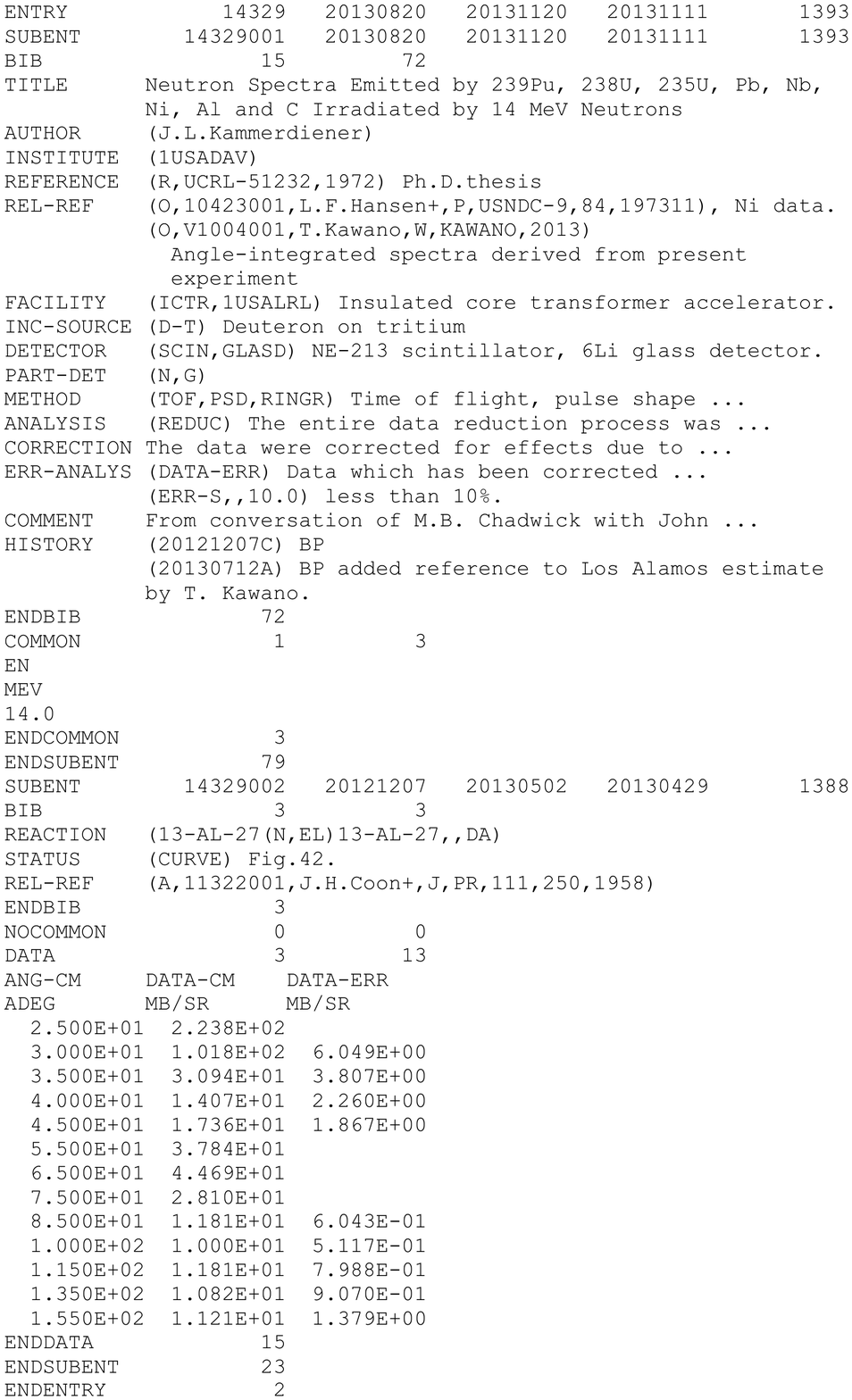}}
\caption{Fragment of  EXFOR entry  {\it 14329} based on the 1972 University of California  Davis thesis by J.L. Kammerdiener  \cite{Kam72}.}
\label{fig:x14329}
\end{figure}

Data recovery (e.g. nuclear archeology)  is one of the important functions of the EXFOR operation. Here, the authors consider  EXFOR entry {\it 14329},  
based on the Ph.D. dissertation of J.L. Kammerdiener \cite{Kam72}, University of California,  Davis.  This thesis represents cost and time-expensive 
measurements with unique target materials: $^{239}$Pu, $^{238}$U, $^{235}$U, Pb, Nb, Ni, Al and C.  These thesis data were preserved as a paper copy only,  
where results were shown as plots.  A new measurement with these targets would require lengthy delay, contingent on a program advisory committee  approval  
process, and  may result in a few million dollars of additional expenses.  
Therefore, this thesis and associated graphic data  were reanalyzed and digitized forty one years later at the National Nuclear Data Center.  
In this endeavor the EXFOR editor \cite{Pik16} and Java  digitizer \cite{Mak13} developed in Sarov, Russian Federation  and Sapporo, Japan, respectively, were used.  
The Fig. \ref{fig:x14329} example illustrates the first 2 of Kammerdiener's 171 thesis data records (subentries). 
The present compilation beginning subentry  contains bibliographical 
and general  information for all subentries, such as 14 MeV incident neutrons energy in the common section,  
while experimental data are compiled in the following subentries. The compilation second subentry reaction field 
(13-AL-27(N,EL)13-AL-27,,DA) indicates elastic neutron scattering differential data with respect to angle (DA), and    
 the subentry status field (CURVE) reveals digitized data.  All experimental data sets are 
placed within DATA and ENDDATA tags, and the original article units are preserved. The well-structured  compilation format 
was intentionally designed for further usage in computer systems. It represents a typical   input file  for the current 
Web and database system that will be discussed in the following section.

\subsection{EXFOR Format}

\begin{figure}[t]
 \centering
 \includegraphics[width=0.95\textwidth]{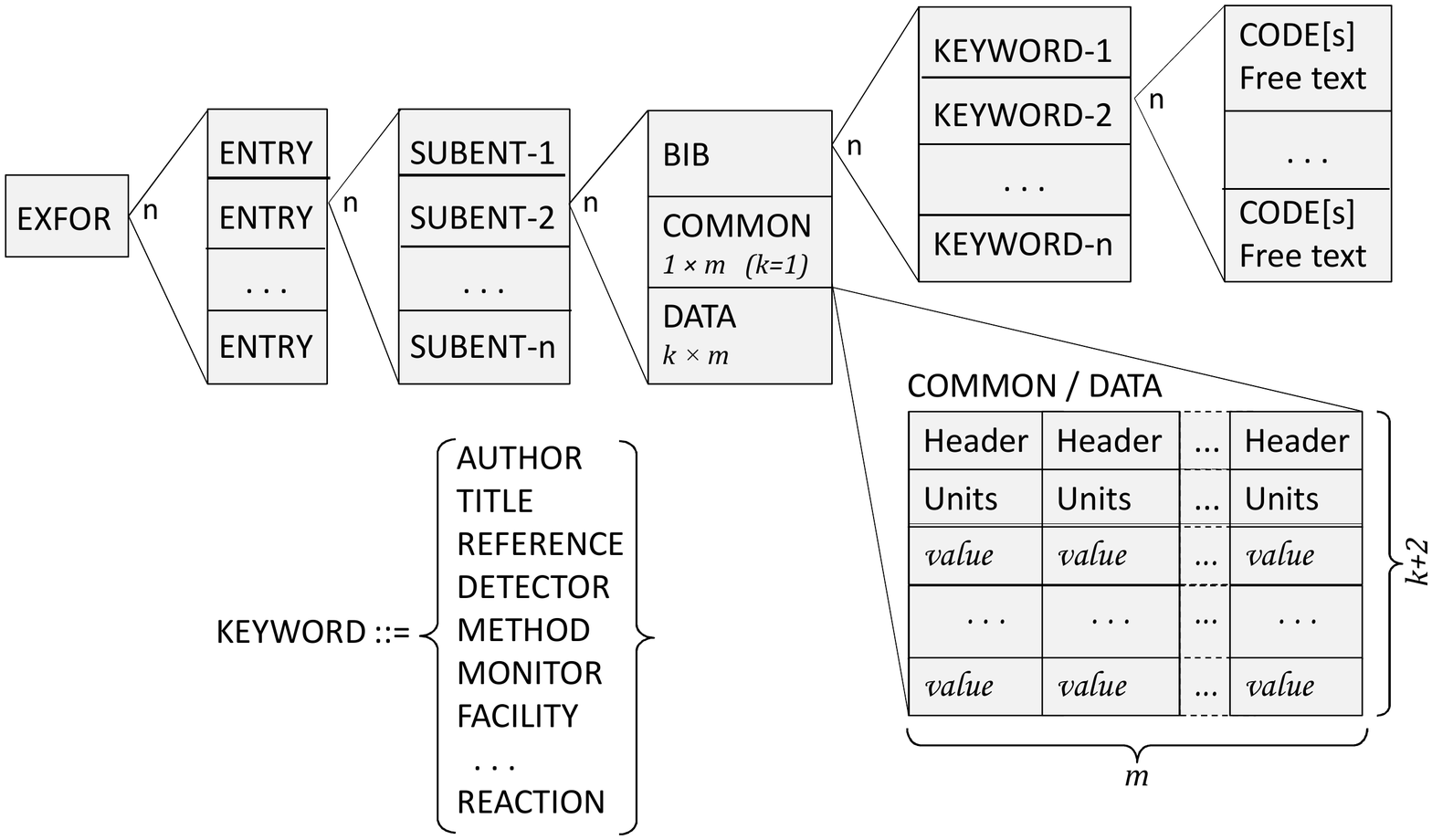}
\caption{Manifold structure of the EXFOR library. Each database entry consists of bibliographical and data subentries. 
}
\label{fig:x4entry}
\end{figure}

The EXFOR database format was introduced for data storage and information exchanges between major data centers as a replacement of the SCISRS format \cite{Holden05}. The present format is 
human-readable,  is 80 characters long, and includes opening and closing tags for all data and metadata sets. 
Efforts to improve the EXFOR format are currently underway \cite{GND}, and these attempts will provide additional avenues for application developments. 
In the present work, the authors will adopt the EXFOR Basics publication \cite{VMclane} as an introductory guide to EXFOR format, for advanced users the LEXFOR manual 
is recommended  \cite{Otto15}.  The EXFOR format is currently used for experimental nuclear data archiving and dissemination,   and its file structure is shown in Fig. \ref{fig:x4entry}. 
The experimental nuclear reaction data library consists of data entries that contain complete records of individual experiments.   
Each experiment may include multiple nuclear reaction data sets (subentries) and several research papers because EXFOR compilers 
group multiple publications from a particular experiment into a single entry. Another important feature of the 
library is that all experimental data are compiled as published, only obvious errors are corrected.

Each entry is assigned a unique accession number; each subentry is assigned a
subaccession number (the accession number plus a subentry number). The subaccession numbers are
associated with a data table throughout the life of the EXFOR system.
The subentries are further divided into the following categories 
\begin{itemize}
\item Bibliographic, descriptive, and bookkeeping information (hereafter called BIB information).
\item Common data that applies to all data throughout the entry or subentry.
\item Data tables.
\end{itemize}
In order to avoid repetition of information that is common to all subentries within an entry or to all
lines within a subentry, information may be associated with an entire entry or with an entire subentry.
To accomplish this, the first subentry of each work contains only information that applies to all other
subentries. Within each subentry, the information common to all lines of the table precedes the
table. 

EXFOR format was originally designed for  data storage and interchange between the four data centers, and it serves well for these purposes. 
However, the compilation format \cite{Schw15} may represent a problem for application developers and other users who are not familiar with format specifics. To overcome this  
obstacle, a simplified computational format (C4) was introduced. The code X4TOC4 \cite{Cullen86}  converts EXFOR data sets to the easy-to-read data columns in C4 format. 
The C4 format is used in the current system for data plotting and further dissemination.

\subsection{EXFOR Database Dictionaries}
Compilation databases often contain an extensive collection of dictionaries.  The EXFOR dictionaries list details for database keywords and codes, and 
their  identification  numbers range between 1 and 999 \cite{Sch14}. 
These dictionaries are periodically updated with new keywords and codes and could be subdivided into three groups:

\begin{itemize}
\item The Archive  dictionaries are the source of the contents of all forms of the dictionaries. 
They   contain   all   keys,   their   expansions,   and additional   codes,   as   well   as   free   text   explanations for compilers.
\item The DANIEL  backup  dictionaries  contain  all  keys,  additional  codes  needed  for  programs, and most of the expansions. These dictionaries are 
designed for interacting with computer programs.
\item The EXFOR  transmission  dictionaries  contain  all  keys  (except  for  Dictionary  26: Unit families \cite{Sch14}),  their  expansions,  and  free  text  
explanations, but not all additional codes needed for the checking and other programs. 
\end{itemize}

\subsection{EXFOR Manual}
EXFOR database, its formats, structures and relevant procedures are extensively described in the manual. Historically, it was maintained 
by the U.S. National Nuclear Data Center, and recently it was transferred to the IAEA, Nuclear Data section \cite{VMclane,Otto15}. The manual 
is absolutely essential for EXFOR database compilations and application developments.

\section{Platform-Independent Data Storage and Dissemination  System}

The development of nuclear reaction data storage and dissemination systems has a very rich history.  
There have been previous efforts to disseminate nuclear reaction data such 
as the Brookhaven Cross Section Information Storage and Retrieval System (CSISRS)  
that was introduced in 1971,  based on the VAX-VMS computer system.  
This dissemination system was a revised version of the original SCISRS. 

The VMS platform became obsolete in the 90s due to the rapid progress of modern computing 
and introduction of World Wide Web technologies. In order to improve the quality of nuclear data services and take advantage 
of the latest software and hardware developments, the NNDC, BNL, and NDS, IAEA, started to work on a  data migration project in 1999, following a 
basic assessment of future options \cite{Tech:01}. This exploratory work culminated at an International Workshop on Relational Database and Java Technologies 
for Nuclear Data, held at BNL, September 11-15, 2000 \cite{NNDCWork:00}. 
As a result of the workshop, the two partners  embarked on a project to migrate its databases to a UNIX-based relational database environment 
and significantly upgrade    Web Services employing the latest computer technologies \cite{Tech:01,Zerkin01}. The migration project was successfully completed 
and the new nuclear data Web service made available beginning in 2004 \cite{pr06}.  General system design and several distinct elements of the system such as 
Nuclear Science References (NSR, {\it http://www.nndc.bnl.gov/nsr}) and Sigma Web Interface ({\it http://www.nndc.bnl.gov/sigma}) have been described previously \cite{pr11,pr08}. 
At the same time other popular features such as the very advanced and constantly evolving experimental nuclear reaction Web and data services were never comprehensively described in the literature. 
This publication is intended to fulfill this task and provide an extensive description of nuclear reaction data services.

In the present work, the authors describe the current status of the EXFOR database project, which creates one of the foundations for the modern nuclear data Web services. 
The current system has been  designed as multiuser and platform independent environment \cite{Zerkin05}.  The cross platform capabilities have been achieved by extensive usage of Java technologies. 
Java codes can be developed on any device, compiled into a standard bytecode and  run on any device equipped with a Java virtual machine.  
The current system includes Java Web servlets, database connectivity and wrapper classes for a ZVView plotting package \cite{ZVView}, and legacy FORTRAN and C codes.  
It is based on commercially available software and hardware and contains many features of enterprise computer portals dealt with on a daily basis.  
Two versions of the EXFOR database and Web retrieval systems are presently distributed:  server and personal computer (portable) for multiple and individual clients, respectively. The server installation is briefly described in Ref. \cite{pr06} while the personal computer version is build on a standard compact disk that is available from the IAEA upon request. 
In subsequent subsections, the authors will describe hardware and software computer environments, followed by software application developments, careful analysis of Web features, and a worldwide impact on nuclear data users.

 \subsection{EXFOR Relational Database}
\label{sec:Scope}
The EXFOR database contains nuclear data sets for $\sim$22,000 experiments. 
Rapid access to these data is of paramount importance for nuclear science and technology professionals.  
The first modern EXFOR Web dissemination and data storage system was developed at the NDS, IAEA, in cooperation with the NNDC, 
BNL \cite{Zerkin05}. It is a multi-platform product that was developed for the nuclear data services migration project 
to a Tomcat Web and Sybase/MySQL database servers environment  \cite{pr06}.  Later, the nuclear databases were transferred from Sybase to   MySQL servers.  
The modern relational database servers allow access to  data using  Structured Query Language (SQL). The SQL language is 
easy to learn and use, it allowing management of the databases and building  Web interfaces for data searches, management, and updates. The underlying software converts these 
requests into corresponding SQL statements that are passed to the database to harvest  database outputs. 
  
MySQL server is a robust database system that can be operated on multiple computer platforms using community or enterprise licenses. 
Initially the NDS, IAEA service was based on MySQL, while the NNDC operated  enterprise Sybase servers and later transferred  to   MySQL enterprise servers.  
In  recent years, both centers migrated to  MySQL community edition servers. 

The EXFOR database has been successfully integrated with the Evaluated Nuclear Data File (ENDF) libraries \cite{endf, Chadwick11,11Kon,15Kon,11Shi,07Zab,11Ge},  
bibliographical Computer Index of Nuclear (reaction) DAta (CINDA) \cite{cinda,cindam}, and Nuclear Science References databases \cite{pr11}. 
Its contents are also used in Sigma Web Interface \cite{pr08}; some of its components are used in JANIS (JAva-based Nuclear data Information System independently developed by the NEA Databank) \cite{NEA}. 
The EXFOR relational database project has evolved and matured substantially in the last fifteen years compared to its initial version \cite{Zerkin05,pr06}, 
and an extensive description of its latest features is presented below.

\subsubsection{EXFOR Database Schema}

The schema describes a database structure in a formal language supported by the database management system. 
In a relational database, the schema defines the tables, fields, relationships, views, indexes,  procedures, and other elements. 

\begin{figure}[]
 \centering
 \includegraphics[width=0.95\textwidth]{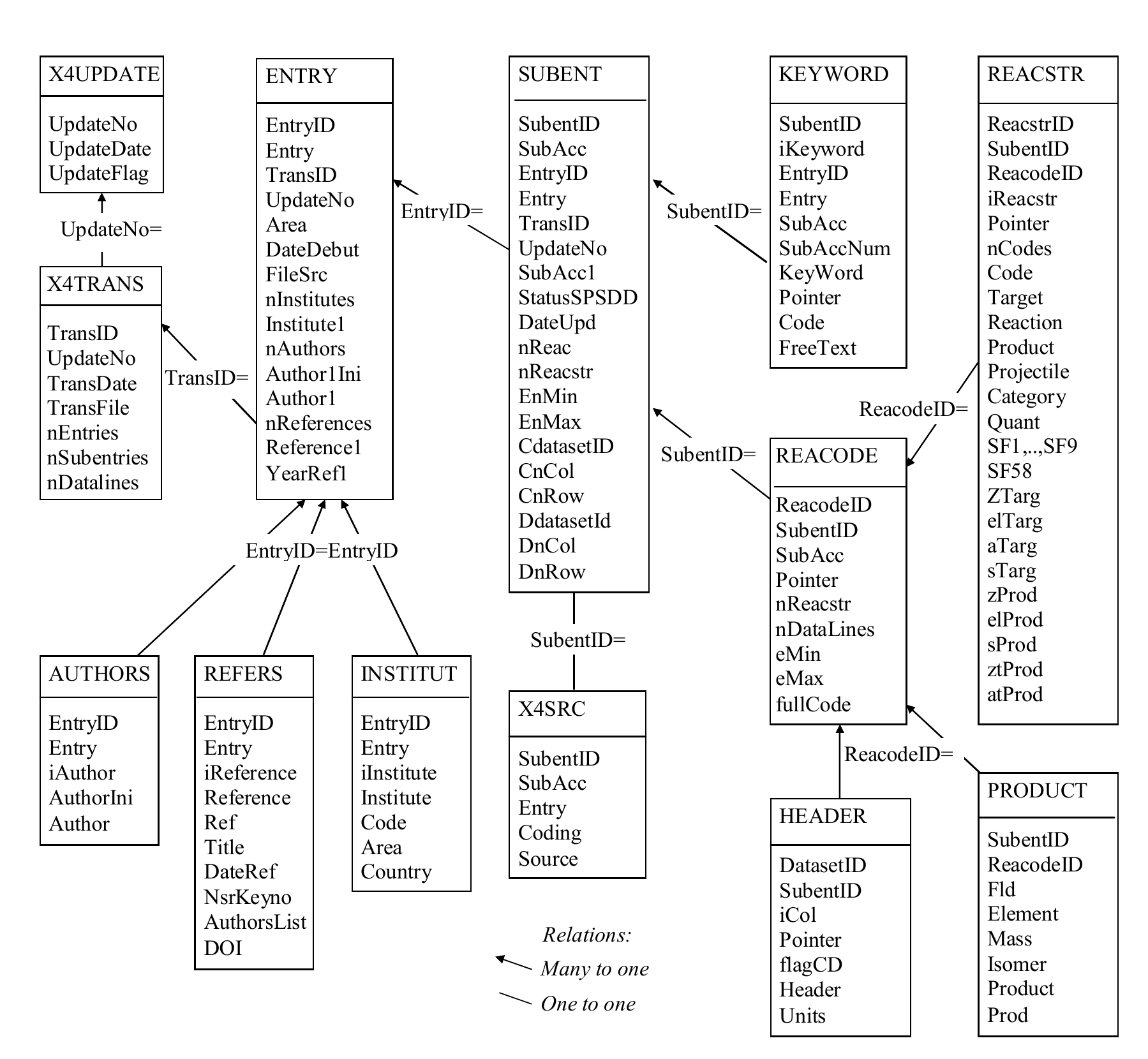}
\caption{The EXFOR database schema.}
\label{fig:x4schema}
\end{figure}

The simplified version of EXFOR relational database schema is shown in Fig. \ref{fig:x4schema}. 
The EXFOR schema includes tables that contain data, metadata, and dictionaries 
such as X4SRC, ENTRIES, and AUTHORS, respectively. It includes relationships between data tables, {\it e.g.} ENTRY and  SUBENT meta data tables 
are linked using an EntryID relationship. The current database schema provides a glimpse into the present system storage of experimental nuclear reaction data sets.

\subsubsection{Database System Operation}

\begin{figure}[]
 \centering
 \includegraphics[width=0.95\textwidth]{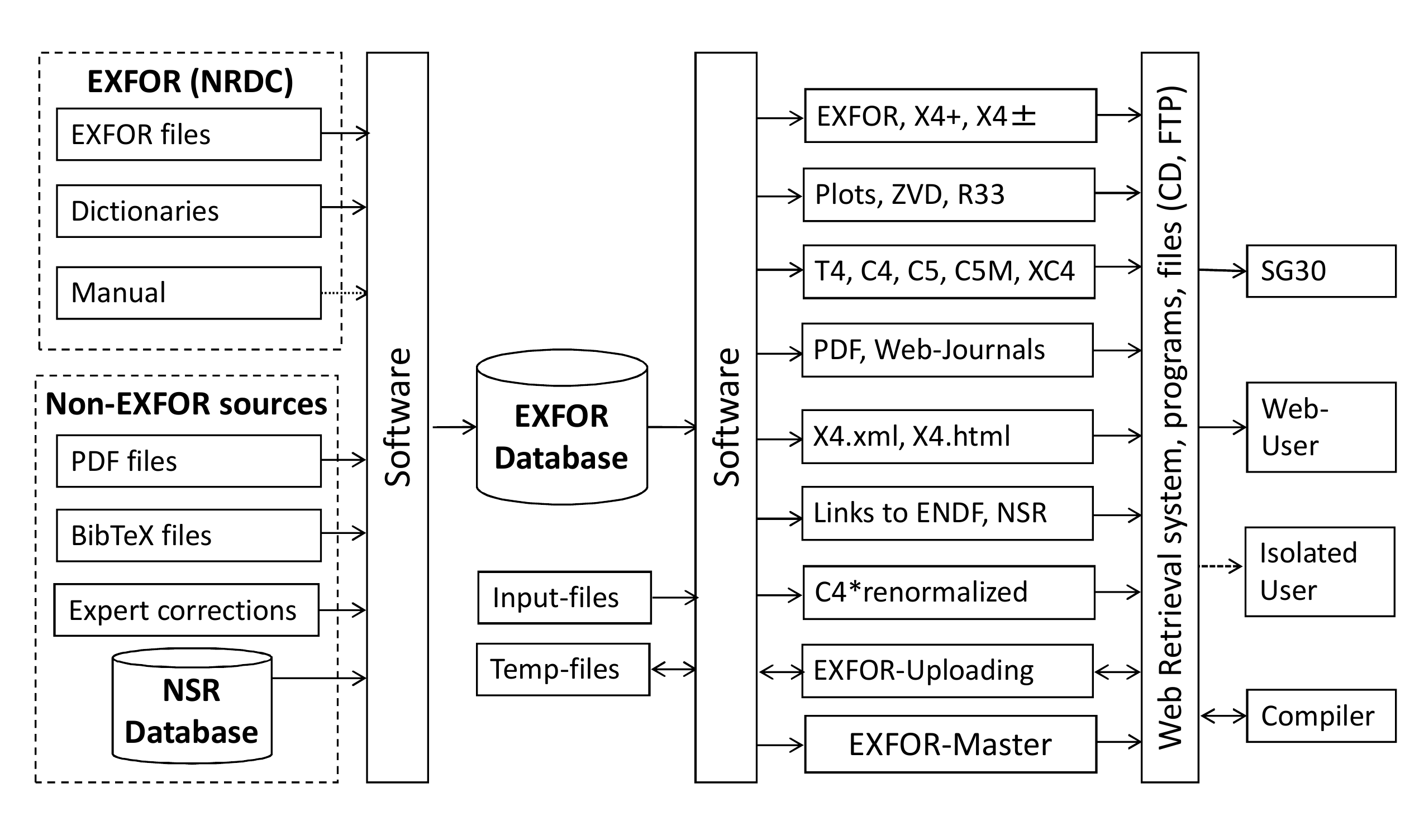}
\caption{The EXFOR  operation workflow.}
\label{fig:x4system}
\end{figure}

The EXFOR operation includes contributions from the data groups affiliated with the network of Nuclear Reaction Data Centres (NRDC) \cite{NRDC} and Non-EXFOR data sources. The NRDC supplies 
the database with new and revised data compilations, dictionaries, and manual updates. Bibliographical metadata, original publication PDF files,  
and data renormalization information are provided through separate channels. All of these nuclear science and 
bibliographical data are processed and stored in a relational database that incorporates text compilation and article/document PDF files for the majority of EXFOR entries. 
Due to major publishers' subscription rules, the PDF files are not available for external users. 
The Java-based computer programs process contributions and fetch data retrievals as EXFOR-formatted files, ZVView plots, PDF files, and many other custom formats for remote Web and local users. 
  The EXFOR database operation workflow diagram is shown in Fig. \ref{fig:x4system}.

\subsection{EXFOR Web Interface}
\label{sec:WEB}
Since introduction of the World Wide Web, it is essential for data and research centers to provide direct access to nuclear databases \cite{pr06,Zer07,Has08,Mom16,NRV17}.  
The EXFOR Web retrieval interface is an integral part of both IAEA and NNDC Web Services \cite{pr06,Zer07}.  
To provide more robust, scalable architecture, satisfy cyber security requirements, and protect nuclear data services
from a single-point failure, the Web  and database servers are physically separated. 
The Web server has the Apache/Tomcat \cite{apache,tomcat} Web production environment installed.  
All servers are running the Linux (Red Hat, SUSE, etc.) operating system. The  initial version of the present  interface became 
operational fifteen years ago and eventually grew to its  current state. The interface is based on current 
computer technologies and provides retrieval of database content in HTML, Text, BibTex, and PDF formats.  It also produces 
specialized outputs in C4/C5,  Application Programing Interface (API)/XML, and R33 for nuclear model codes, application developers, and
Ion Beam Analysis Nuclear Data Library (IBANDL) \cite{ibandl} users community, respectively.

Data plotting is  performed using the ZVView software package  that has been written in C and designed for evaluators of 
nuclear reaction data to perform efficient interactive analyses of cross section data retrieved from the nuclear data  libraries \cite{ZVView}. 
The main function of ZVView is to plot these data for  comparison, using a variety of options to study graphical, 
numerical, and bibliographic information along with the possibility of analyzing the results of the user's evaluations. 
Since 2010, the ZVView program has been extended to plot 2-dimensional arrays Z(X,Y) with the main purpose of displaying correlation matrices. 
The package allows users to change plotting parameters, such as: type of data view (curve, histogram, map, 3D, animation, symbols of points, colors, error bars), 
logarithmic and linear scales, zoom, split plot to sub-windows, smoothing by least-squares method, choice of data and authors to be plotted, 
 scans of their points, and changing language on the fly. The Web-ZVView program implements the majority of these functions as integrated part 
of EXFOR and ENDF Web retrieval systems.

The current Web interface is a very sophisticated product that reflects the overall complexity of nuclear reaction data. 
The immense variety and volume of the archived quantities require advanced data retrieval capabilities;  
at the same time a large number  of EXFOR users are still looking 
for simple retrievals of reaction cross sections. To accommodate these different user groups,  four levels of EXFOR retrievals have been introduced: 
basic (simple), extended, keywords, and expert.

\subsubsection{Simple EXFOR retrievals}
\label{sec:Retrievals}
\begin{figure}[]
 \centering
\includegraphics[width=0.7\textwidth]{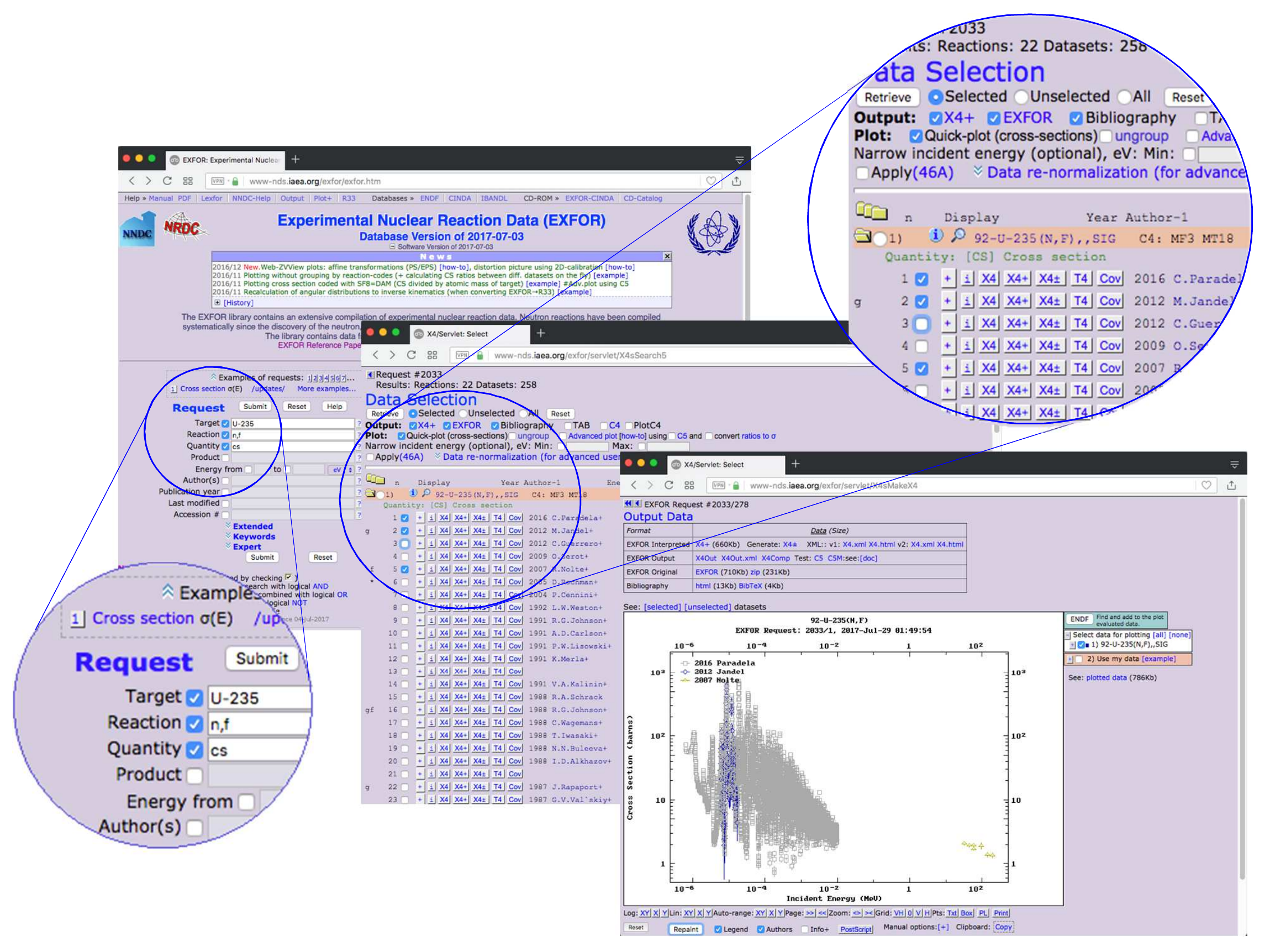}
\caption{An example of simple cross section retrieval for $^{235}$U(n,f) reaction, data inputs for the EXFOR front and data selection pages were magnified. 
The EXFOR Web Interface is publicly available at the websites of the   Nuclear Data Section, {\it http://www-nds.iaea.org/exfor},  
 the National Nuclear Data Center, {\it http://www.nndc.bnl.gov/exfor}.}
\label{fig:EXFOR}
\end{figure}

The majority of EXFOR users are not familiar with the database's advanced structure or its philosophy. 
This user group includes nuclear scientists, engineers, and graduate students that  need direct access 
to the latest nuclear  reaction data sets. To satisfy their needs, the authors have developed simple EXFOR retrievals 
that are based on target, reaction, experimental quantity, product, energy range, authors, publication year, and accession number search parameters. 
All search fields are based on a logical ``OR" operator ($\parallel$) that returns the Boolean value true if either of the operands is true and returns false otherwise.  
Therefore empty fields are ignored by a data search query, and users can easily construct database queries using  Web interface search fields. 
To demonstrate the present Web interface graphic and data capabilities the authors will consider a case of $^{235}$U(n,f) reaction cross sections.  
Fig. \ref{fig:EXFOR} shows a simple retrieval for the following input parameters:  target = $^{235}$U, reaction = n,f, and  quantity = cs. 

To improve user experience, a simple text string Google-like search has been recently added to the EXFOR Web interface. 
It is located on the top right side of the EXFOR front page. This search is intended for less sophisticated EXFOR users who 
prefer to mine experimental data using  
common nuclear physics concepts.   The EXFOR interface also contains online help displayed as question mark buttons on the right side of the search boxes. 
These buttons produce pop-up windows that have information on text box input parameters and provide access to EXFOR dictionaries. 

\subsubsection{Advanced EXFOR retrievals}

EXFOR is the only nuclear physics database that stores extensive descriptions of experimental results and facilities. In fact, 
these data represent a treasure trove for investigation of research trends in nuclear physics \cite{Pritychenko15} and 
provide complementary background for nuclear data evaluations. These  data can be extracted using extended, keywords, and  expert requests.  

The extended retrieval provides additional capabilities for creating sophisticated database queries that include bibliographical metadata, 
geographical origin (area, country, institute), NSR keynumbers (NSR database analogues of EXFOR accession numbers), and compilation timing information. 

Keyword retrieval allows queries for experimental quantities  that cannot be easily found anywhere else: 
detector, method, monitor, facility, incoming source and spectrum, data analysis, detected particles and radiations, and experimental data origins.

The expert retrieval supplies capabilities for searching for outgoing particle parameters, data units, ranges for experimental data points and 
EXFOR compilation centers, individual compilers, and data transmissions. As it was mentioned previously, all text box search fields are 
implemented as a logical ``OR" operator and, therefore, missing search fields are ignored by SQL data queries.

\subsection{EXFOR Examples of Requests and Video Guides}

The EXFOR system has matured over the years into a very advanced product and several complex features  represent a challenge 
for new users. To  overcome these issues,  the webpage News display box, numerical and plotting examples of requests, and video guides ({\it https://www-nds.iaea.org/exfor/x4guide/}) 
have been prepared. These resources provide constant updates on new EXFOR system features and share 
best user practices. The EXFOR examples are intended for user education purposes and supply 
additional insights on interface capabilities. The broad scope of examples helped to improve EXFOR user experiences,  and 
such policies contributed  to the worldwide popularity of  the  service.

\subsection{Worldwide Impact of EXFOR Web Interface}

The EXFOR interface is a gateway to the latest experimental nuclear reaction data stored at the IAEA and NNDC databases,  
and it has a very broad spectrum of usage worldwide. In order to provide the best quality of services the authors monitor the interface operation \cite{pr06,pr12}. 
Monitoring includes collection of web statistics, geographical coverage, network data analysis,  and estimates of database usage. 
The EXFOR database retrieval statistics are very conservative and completely based on a count of successful database retrievals;  
any Web browser hits are ignored. Retrievals for non-existing entities in the 
database (e.g. author, reaction, target), which produce an empty output file are rejected. Complete information for 
each retrieval is recorded in separate Web statistic data files, and usually,  $\sim$300 retrievals/day are observed  
from the IAEA and NNDC websites. 

\begin{figure}[]
 \centering
 \includegraphics[width=0.7\textwidth]{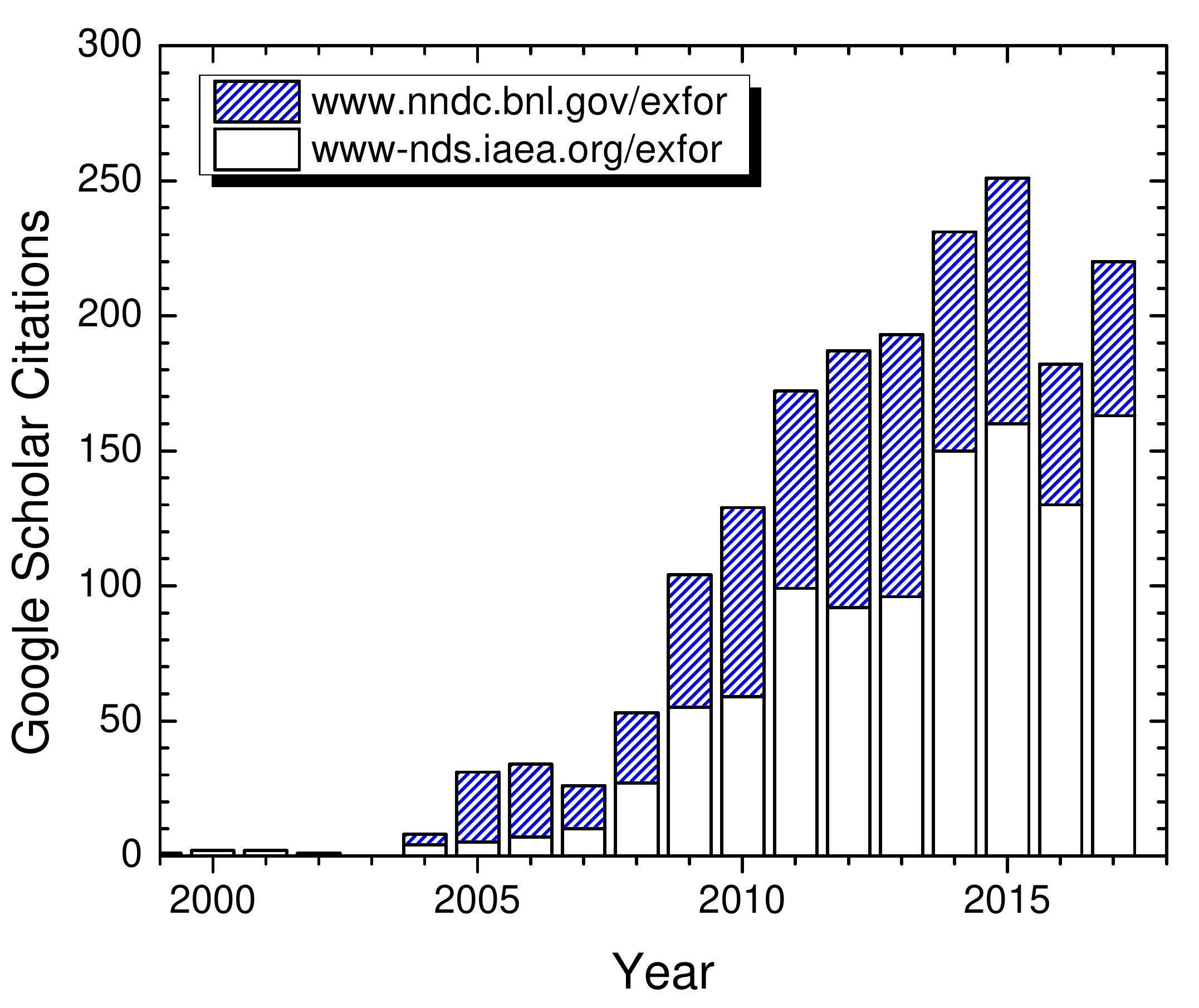}
\caption{The worldwide impact of EXFOR Web application based on the NDS, IAEA, and NNDC, BNL citation statistics. The plot shows an incremental growth of website Google Scholar 
citations over the years.}
\label{fig:GScholar}
\end{figure}

The absolute number of Web retrievals provides only a partial estimate of  product usage. To further estimate the impact of the current product, 
retrieval statistics can be complimented with the Google Scholar data. The Scholar Web application allows retrieval of   
information on the EXFOR Web application usage from research articles. A  simple Google Scholar search  
for the NDS {\it http://www-nds.iaea.org/exfor} and   NNDC {\it http://www.nndc.bnl.gov/exfor} text strings results in $\sim$1,100 and 
$\sim$770 citations, respectively. These citations originate from  research papers where the EXFOR interface address was mentioned.    
The  growth curve shown in Fig. \ref{fig:GScholar} is well matched with the Web retrieval statistics \cite{pr12}. 
The agreement between these trends  provides a verification of the present analysis 
and confirms a broad usage of the EXFOR Web interface.

\begin{figure}[]
 \centering
 \includegraphics[width=0.8\textwidth]{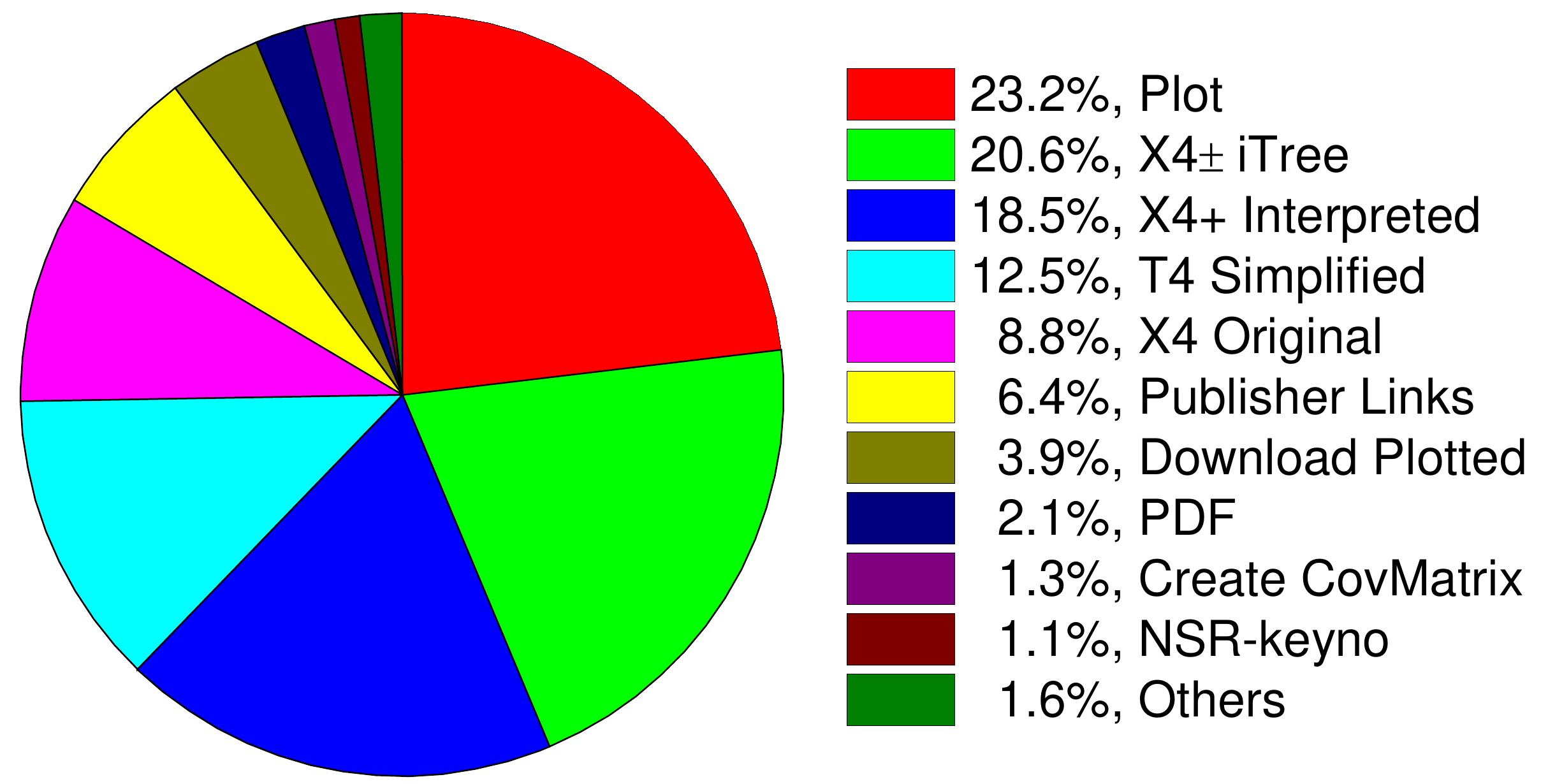}
\caption{Google Analytics statistical analysis for  EXFOR Web interface. 
}
\label{fig:Pattern}
\end{figure}

Complementary insights on EXFOR usage can be obtained from  the  Web interface usage pattern pie chart shown in Fig. \ref{fig:Pattern}.  
The chart is based on a Google Analytics analysis of NNDC and NDS Web traffic $\sim$100,000 interactions per year (data searches are not included) 
and $\sim$2,500 unique users per month. The EXFOR plot, X4$\pm$ interactive tree (iTree), X4+ interpreted, T4 simplified and original 
format retrievals constitute 23.2, 20.6, 18.5, 12.5, and 8.8$\%$ of total operations, respectively. Other noticeable contributions 
include access of publisher Digital Object Identifier (DOI) links  and NSR database, downloads of plotted data and PDF files, 
and covariance matrix calculation using experimental uncertainties. Usage of the retrieval system can be summarized as follows:
\begin{itemize}
\item	EXFOR data  downloads in various interpreted formats (iTree, X4+, T4 -  $\sim$50$\%$). 
	These data transfers exceed the original EXFOR format retrievals ($<$9$\%$).
\item	EXFOR plotting with the subsequent download of plotted data as text or html tables ($\sim$27$\%$) and construction of covariance matrices using experimental uncertainties ($\sim$1$\%$).
\item	DOI access of publisher web sites (6.4$\%$) and NSR web retrieval system ($\sim$1$\%$) and download of PDF copies ($\sim$2$\%$).  The last option is available for authorized users only due to copyright restrictions.
\item	Other tools and operations ($<$2$\%$); for example, extremely valuable for nuclear physicist or engineer  EXFOR cross section renormalization  
(fifty-two unique clients in 2016 represent $\sim$0.1$\%$ of all operations).
\end{itemize}

\section{Related Nuclear Reaction Data Projects}

The EXFOR  computer database is an integral part of nuclear data Web services, is firmly related to  nuclear reaction data evaluations, and requires access to nuclear bibliography. 
The combined development of the experimental, evaluated, and bibliographical  nuclear reaction databases was a cost effective solution that allowed  reuse of Java classes, 
Web implementation of plotting packages, and reduction of overall time for application development and testing.

\subsection{ENDF}

 ENDF is a core nuclear reaction database containing evaluated (recommended) cross sections, 
spectra, angular distributions, fission product yields, thermal neutron scattering, and photo-atomic and other data, 
with an emphasis on neutron-induced reactions. ENDF library evaluations are based on theoretical calculations normalized to experimental data  
with an exception of neutron resonance region, where priority is given to experimental data. 
ENDF evaluations consist of complete neutron reaction data sets for particular isotope or element (ENDF material). They cover all neutron reaction channels 
within the 10$^{-5}$ eV - 20 MeV energy range.  Nowadays, in many evaluated data libraries, this energy range is extended to higher energies. 
All these data sets are stored in the internationally-adopted ENDF-6 format that is maintained by the Cross Section Evaluation Working Group (CSEWG) 
and publicly available at the NNDC and NDS websites, {\it http://www.nndc.bnl.gov/endf} and {\it http://www-nds.iaea.org/endf}, respectively.  
Cross section evaluations are of great importance for 
nuclear energy and science applications, and they are performed internationally for ENDF, 
JEFF, TENDL, JENDL, ROSFOND  and CENDL   libraries \cite{Chadwick11,11Kon,15Kon,11Shi,07Zab,11Ge} in the USA, Europe, Japan, Russian Federation, and China, respectively.

The EXFOR project is a primary source of experimental nuclear reaction data for  neutron cross section evaluations. 
In practice, many ENDF evaluations are produced using  nuclear reaction model codes EMPIRE or TALYS \cite{2007EMPIRE,2012TALYS} 
that contain EXFOR database connectivity modules. The computer codes are used for calculation neutron-induced reaction theoretical curves, 
data analysis, and comparison with experimental data.  
These evaluations are critically important for many applications because they provide evaluated neutron 
cross sections for operation of nuclear power plants and GEANT and MCNP computer codes \cite{GEANT,MCNP}.

\begin{figure}[]
 \centering
\includegraphics[width=0.7\textwidth]{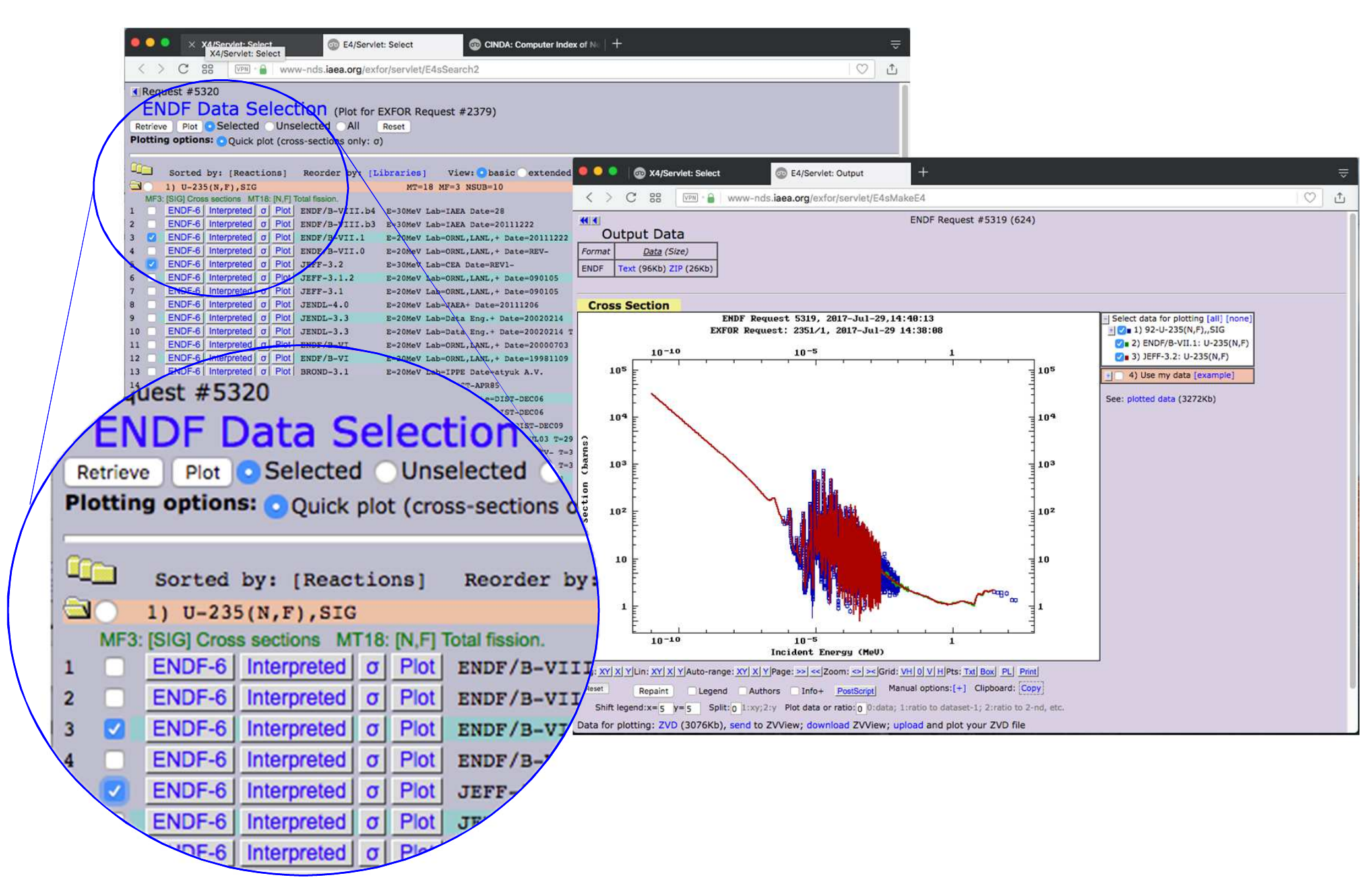} 
\caption{ENDF/B-VII.1 and JEFF-3.2 evaluated and experimental cross sections for $^{235}$U(n,f) reaction, data inputs for the ENDF data selection pages were magnified. 
The visualized evaluated data have been processed with the PREPRO code \cite{Prepro} at T=293.16 K. ENDF Web Interface is publicly available 
at the websites of the   Nuclear Data Section, {\it http://www-nds.iaea.org/endf},  
 the National Nuclear Data Center, {\it http://www.nndc.bnl.gov/endf}.}
\label{fig:ENDF}
\end{figure}

The ENDF Web interface takes its contents from a separate ENDF relational database. It contains the most complete 
international collection of evaluated nuclear reaction libraries, includes data analysis capabilities and  integrated with experimental data. 
The  interface  shown in Fig. \ref{fig:ENDF} serves evaluated data in many output formats and incorporates the ZVView program \cite{ZVView}.
The IAEA and NNDC interfaces are slightly different on a client side  and completely identical on the server side. 
These differences have been introduced to  reflect different user requirements. Both interfaces allow users to 
search a large number of evaluated libraries for relevant evaluations using target, reaction, quantity, product, and other parameters.   
For example, consider ENDF/B-VII.1 and JEFF-3.2 evaluated nuclear data retrieval for $^{235}$U(n,f) reaction cross sections in conjunction with experimental data: 
library = ENDF/B-VII.1, JEFF-3.2, target = $^{235}$U, reaction = n,f and  quantity = cs. 
This retrieval can be accomplished from the main ENDF interface ({\it http://www.nndc.bnl.gov/endf}) or, as shown here, by clicking an ENDF button next to the  experimental cross sections plotted in Fig. \ref{fig:EXFOR}. 
The ENDF database output demonstrates that ENDF/B-VII.1 and JEFF-3.2 evaluated cross sections agree well with experimental values. 

\subsection{CINDA}

Nuclear bibliography is essential for nuclear reaction data compilations. CINDA contains many historic publications 
 that are not available in any other database. These publications are grouped by experiment, and the database allows application of energy range criteria for  data searches. These 
 unique capabilities are still in demand by nuclear data evaluators. Currently  separate CINDA data compilations  are terminated,  
and CINDA contents are extended  by new information from  the EXFOR and NSR databases.  
 
\begin{figure}[]
 \centering
\includegraphics[width=0.6\textwidth]{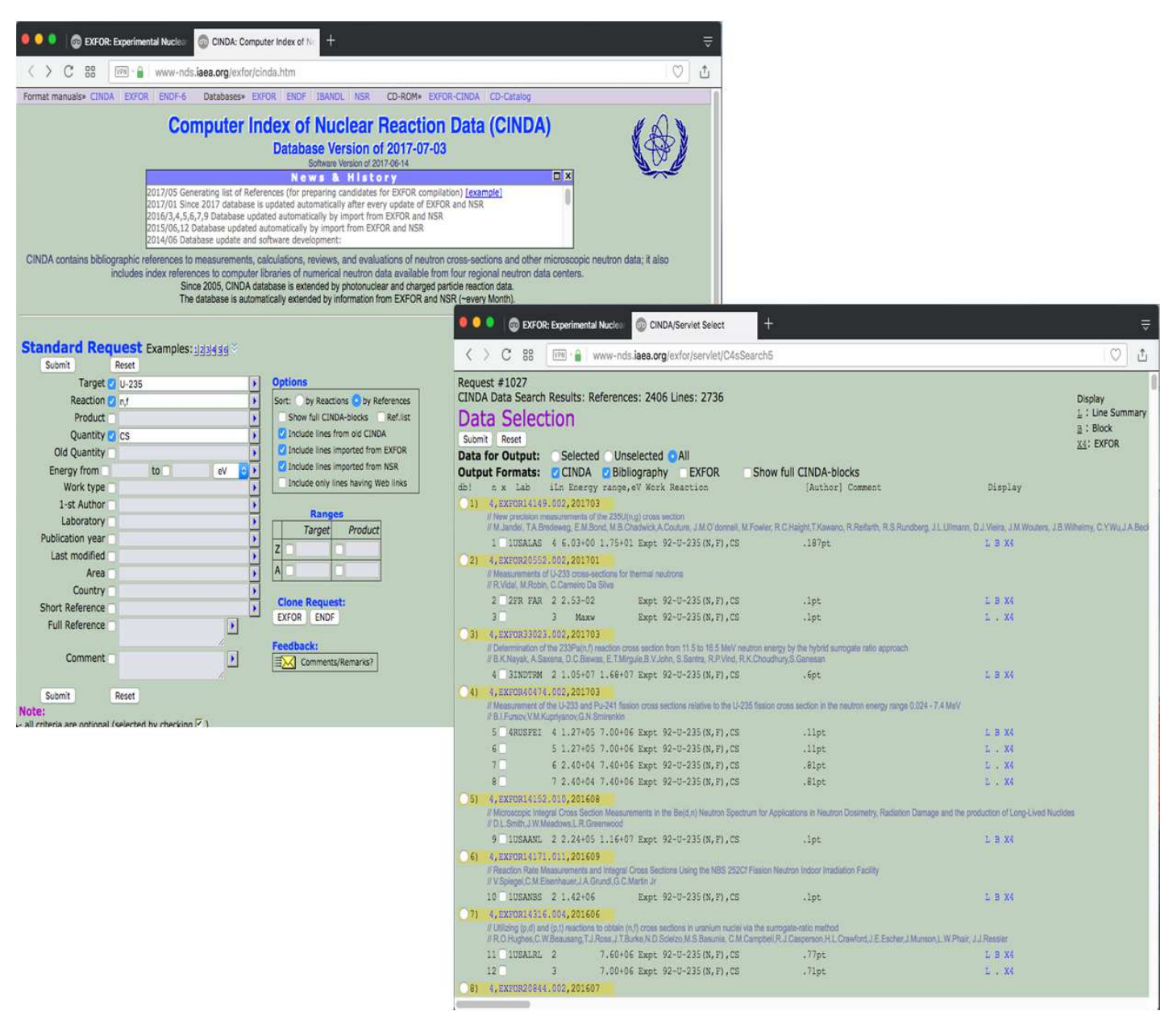} 
\caption{Nuclear bibliography for $^{235}$U(n,f) reaction experimental cross sections. The CINDA Web Interface is publicly available at the websites of the   Nuclear Data Section {\it http://www-nds.iaea.org/cinda},  
 the National Nuclear Data Center {\it http://www.nndc.bnl.gov/cinda}.}
\label{fig:CINDA}
\end{figure}

 CINDA contains the most extensive nuclear reaction bibliography;  
 it is integrated with the EXFOR Web retrieval system and allows users  to search for information that is not available in EXFOR. 
 The database interface  includes all major EXFOR search fields and provides data outputs in text and HTML formats. For example, to retrieve 
 nuclear bibliography for $^{235}$U(n,f) reaction cross sections the following input parameters are assumed: target = $^{235}$U, reaction = n,f and quantity = cs. 
 As shown in Fig. \ref{fig:CINDA}, CINDA database output consists of 2,406 publications and 2,736 reactions. 

\section{New Features of EXFOR Web Interface}
\label{sec:Apps}
Originally, EXFOR Web Interface was created to perform only basic operations: data search, retrieval, and presentation. It is now part of a large and developed Web System including extended functionality 
and tools for regular [external or traditional] users, EXFOR compilers, and ENDF and ENSDF \cite{ensdf} evaluators. 
There are many interesting EXFOR Web interface features that have been developed in recent years, and  readers may explore them 
using the News display box  information and  the search textbox links displayed above. The next section will concentrate  
on  experimental data processing, renormalization, covariance matrix,  and inverse reaction cross section calculations. 

\subsection{User Data Uploads}

EXFOR library compilations are conducted by the NRDC compilers in  cooperation with  researchers. The compilers 
often interact with the scientists in order to obtain original data and complementary details; they are required to 
discuss EXFOR compilations before submitting them to the database. On the other hand, scientists often need to compare their latest findings with the existing data sets. 
Consequently, a direct link between  research and compilation activities is necessary. 
It can be accomplished with a simple Email interaction using the {\it http://www.nndc.bnl.gov/exfor/x4submit.htm} webpage for guidance 
or using a sophisticated Web tool interface: {\it https://www-nds.iaea.org/exfor/x4data.htm}. The Web tool system does not require a password, however, 
users have to authenticate themselves with an  ``I'm not a robot" test. 
On the following page, scientists can provide author, reaction, method information, and data description, and copy and paste their data 
in x, y, $\delta$y, or user-specified formats into the corresponding text box. Upon completion of this operation  
data and associated metadata are converted to EXFOR format and sent to EXFOR Web retrieval system for data operations 
such as plotting, calculations, and comparisons with existing data from the EXFOR and ENDF databases. Users can process their data by utilizing   
 Web tools for constructing covariance matrices or inverse reaction cross section calculations. 
The interface also can be used for new data submissions to the NRDC network for further analysis and  compilation. 

The complementary general purpose myPlot package provides additional avenues for uploading and displaying Web interface users' data.  
This Web-ZVView program extension is available under {\it https://www-nds.iaea.org/exfor/myplot.htm}.

\subsection{Data Renormalization and Expert Corrections}

Experimental results are never perfect; they rather represent the best effort of a particular group of authors and emulate the overall understanding 
of nuclear physics processes and technology development at the time. In present days, nuclear scientists and data evaluators have a better grasp of nuclear physics processes and limitations of 
experimental techniques. Besides, many results in the past were produced relative to presently-outdated nuclear standards (monitor measurements); these standards have improved over time,    
and  data reanalysis and corrections are urgently needed.

\begin{figure}[t]
 \centering
 \includegraphics[width=0.65\textwidth]{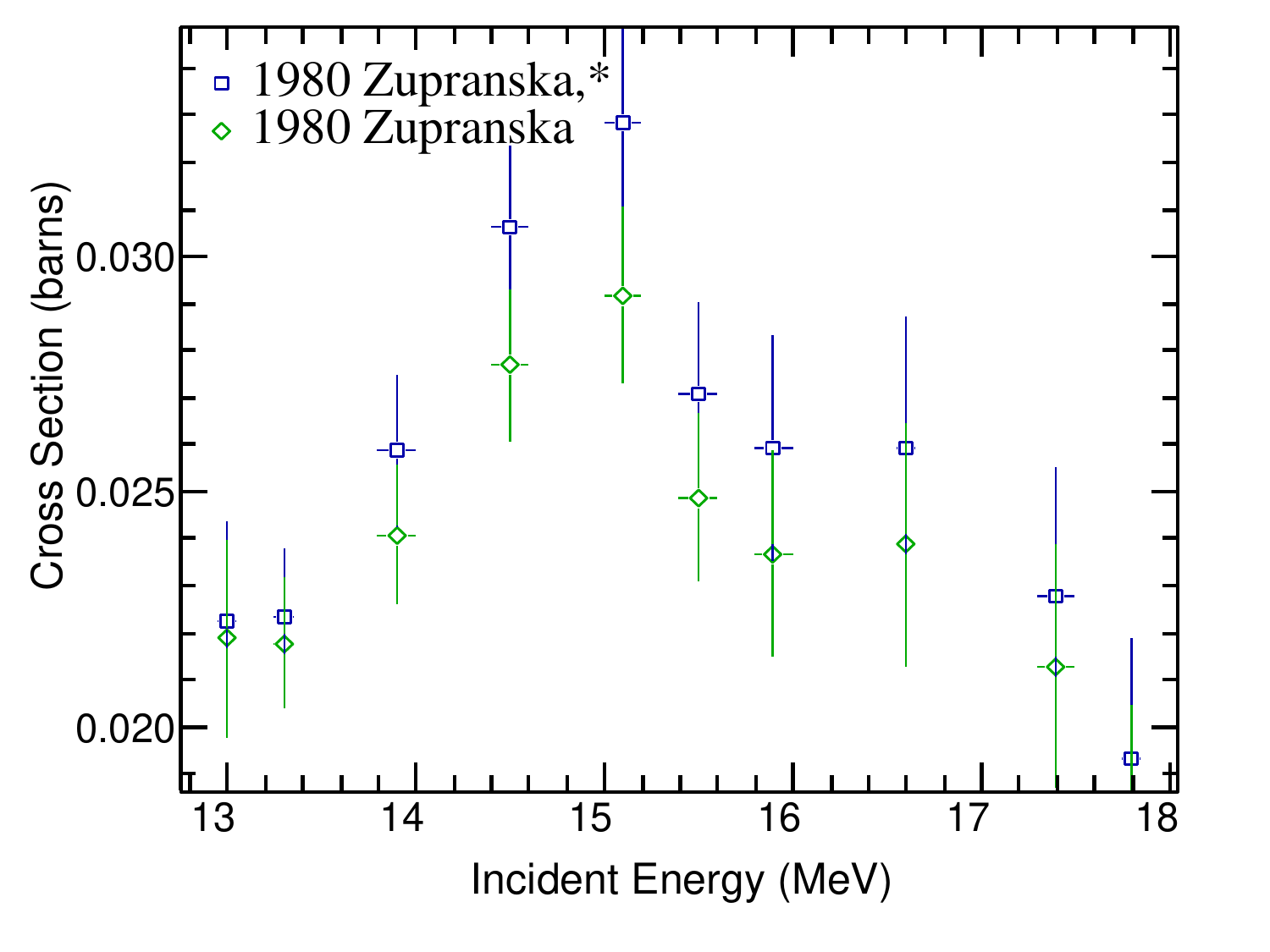}
\caption{Automatic EXFOR data renormalization for the 25-MN-55(n,$\alpha$)23-V-52 reaction data of E. Zupranska et al. \cite{Zup80}. 
The original  and renormalized data are shown as rhomb and square symbols, respectively.}
\label{fig:x4renorm}
\end{figure}

The EXFOR interface addresses these issues with a set of options: automatic data renormalization to new monitor data \cite{Zerkin12} or  
 user and expert recommended corrections \cite{ZolCor}.  The automatic renormalization is available for EXFOR compilations that include information on monitor reaction data. 
An example of automatic EXFOR data renormalization   is shown in Fig. \ref{fig:x4renorm}. 
Here, the  25-MN-55(n,$\alpha$)23-V-52 reaction data and their uncertainties \cite{Zup80} were renormalized  using the latest monitor data  \cite{Carl09,ZolCor} with energy dependent correction factor varying from 1.016 to 1.125.   
Additional details on this subject can be obtained from the EXFOR examples and video guide  ({\it http://www.nndc.bnl.gov/exfor/x4guide/x4renorm2/index.htm}).

\subsection{Covariance Matrix Calculation using Experimental Uncertainties}

\begin{figure}[t]
 \centering
 \includegraphics[width=0.65\textwidth]{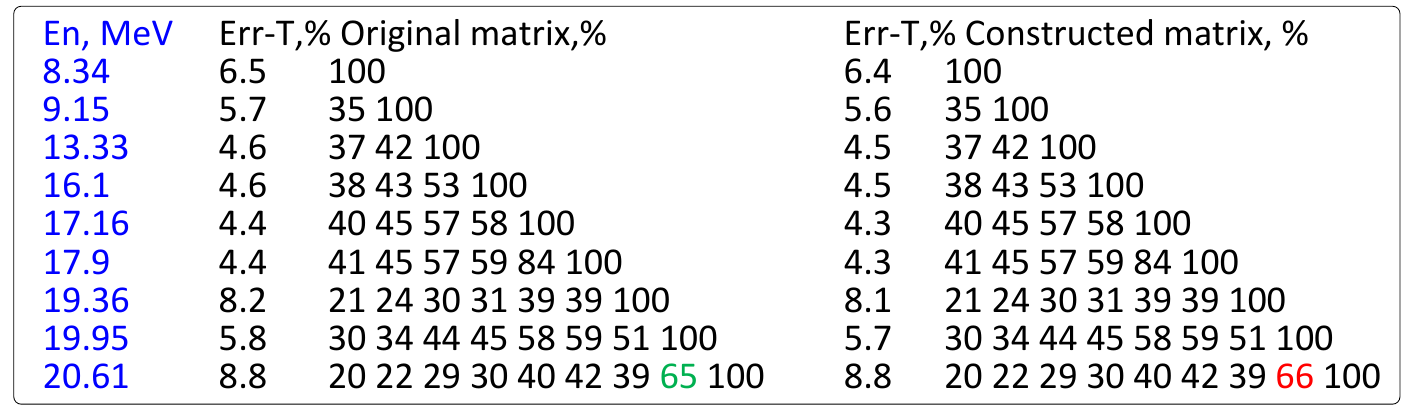}
\caption{Original (left) and Web-constructed correlation (right) matrices. 
Matrix elements were calculated by the Web tool \cite{Zerkin121} using the C. Sage et al. prescription  \cite{Sage10}.}
\label{fig:covar}
\end{figure}

During the last few years, there is a growing interest in covariance data among 
the nuclear physics community.  The experimental nuclear data covariances are needed by 
data evaluators, nuclear model code developers and experimentalists. The  EXFOR database contains  a limited number of  covariances 
and extensive compilations of total, statistical and systematical uncertainties. A recently-developed EXFOR Web interface tool \cite{Zerkin121}  solves this problem by 
interactive calculations of covariance matrix elements, using experimental uncertainties. An example of covariance (correlation) matrix calculation is presented in Fig. \ref{fig:covar}.

\subsection{Inverse Reaction Calculation}
 The EXFOR Web interface collection of data evaluation tools has been recently complimented with 
 an inverse reaction calculator which is based on the nuclear reaction reciprocity theorem \cite{88Rol,08Lar}. 
 It allows extracting a reaction cross section if an inverse reaction is known. 
For 1 + 2 $\rightarrow$ 3 + 4 and 3 + 4 $\rightarrow$ 1 + 2 processes the cross section ratio is 
\begin{equation}
\label{myeq.r6}
\frac{\sigma_{34}}{\sigma_{12}} = \frac{m_3 m_4 E_{34}(2 J_3 + 1) (2 J_4 + 1)(1 + \delta_{12})}{m_1 m_2 E_{12}(2 J_1 + 1) (2 J_1 + 1)(1 + \delta_{34})},
\end{equation}
where $E_{12}$,  $E_{34}$ are kinetic energies in the c.m. system, J is angular momentum, and $\delta_{12}$=$\delta_{34}$=0. 

\begin{figure}[]
 \centering
 \includegraphics[width=0.7\textwidth]{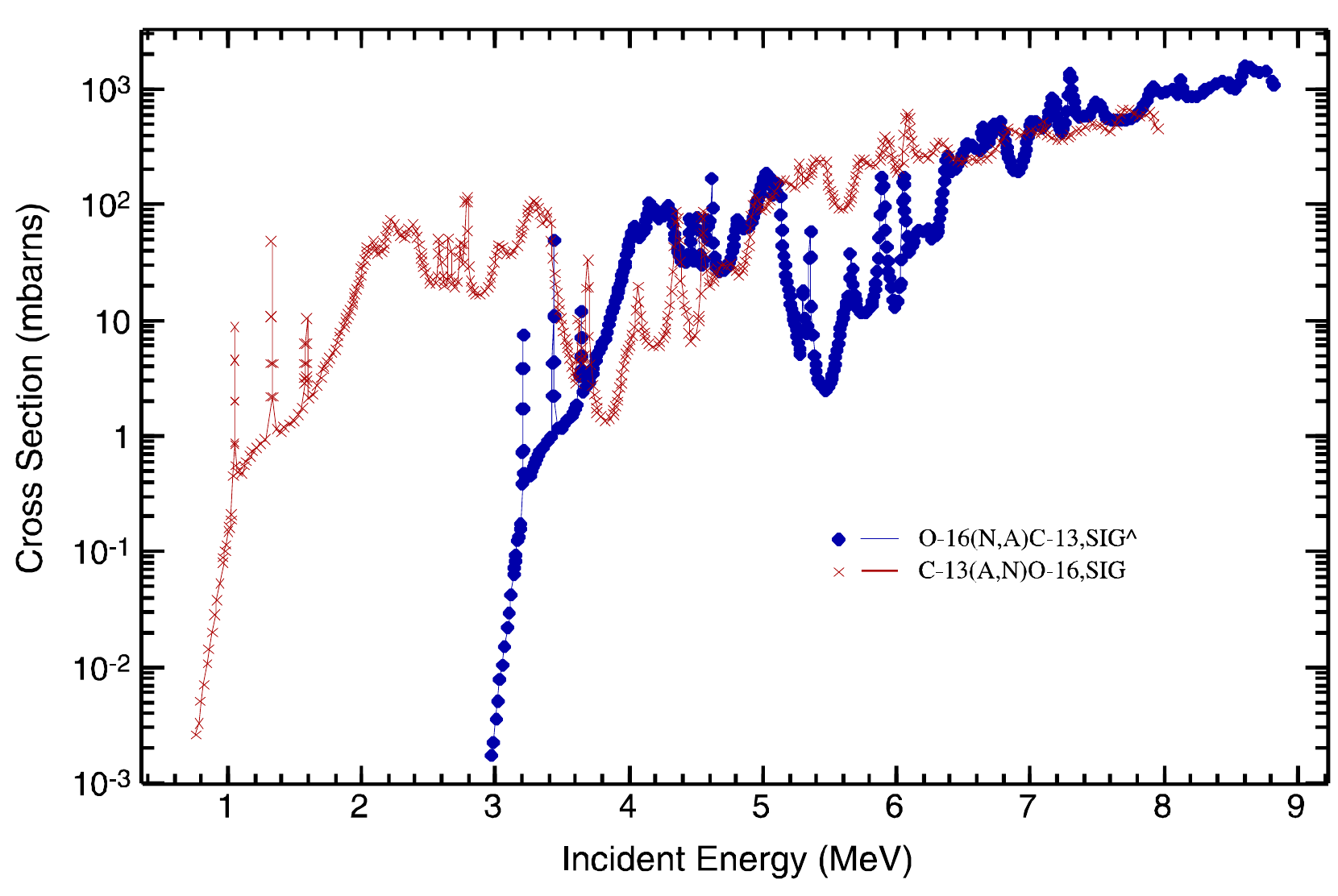}
\caption{Results of $^{16}O + n$  reaction cross section calculation using the $^{13}C + \alpha$ reaction measurement of S. Harissopulos et al. \cite{Har05}. 
The direct and inverse reaction cross sections are shown as red crosses and blue rhombs, respectively. }
\label{fig:16O}
\end{figure}

As an example of application of nuclear astrophysics reaction data for calculations of reactor data,  
one of the major neutron sources for $s$-process nucleosynthesis is $^{13}C + \alpha \rightarrow ^{16}O + n$ reaction,  and its inverse reaction is  
one of the major nuclear reactor neutron poisons. The alpha-induced reactions have been extensively studied due to general availability of 
helium beams, while the inverse neutron reaction measurements are more challenging. Therefore, it is wise to use the well-known alpha-induced reaction on 
carbon for the oxygen neutron capture cross section estimates. 
Using the recent measurement of $^{13}C + \alpha$ reaction  by S. Harissopulos et al. \cite{Har05}  for inverse reaction cross section calculations,   
the corresponding Web interface inputs are: target = $^{13}$C, reaction = a,n, quantity = cs and product = $^{16}$O. 
On the next screen, the example use the S. Harissopulos et al. data set, marking the ``invert data to reaction 8-O-16(N,A)6-C-13,,SIG" checkbox.   
Finally, the calculated reaction cross section sections for the $^{16}O + n$  reaction are shown in Fig. \ref {fig:16O}. 
Such calculations can be performed on all EXFOR data sets or user submitted data sets where the reciprocity theorem is applicable. 
The present tool also allows calculations of angular distributions for inverse reactions.

\section{Conclusion and Outlook}
\label{sec:Conclusions}

EXFOR is the major nuclear physics database for low- and intermediate-energy nuclear reactions. 
The  database and its Web interface are publicly available  through  the IAEA,  ({\it http://www-nds.iaea.org/exfor}), 
BNL, ({\it http://www.nndc.bnl.gov/exfor}),  and mirror websites. The Web interfaces provide transparent and easy access to nuclear reaction and associated  bibliographic information with direct links to the original data sets and articles, where possible. 
Recent application developments  include many features for nuclear scientists and engineers, such as user-friendly data uploads, data renormalization, covariance matrix,  and inverse reaction calculations. 
The authors will continue to address the applied and fundamental science user needs by developing new data access, processing, dissemination, and analysis capabilities.

Further implementation of the advanced analysis features and integration with ENDF and NSR databases  improves database completeness, extends EXFOR scope, 
and ensures the high quality of its contents.  A complementary Google Scholar citation data analysis shows a broad impact of the current interface on nuclear energy and physics research worldwide.  
The EXFOR database and its Web interface represent a robust system that has evolved over the last fifteen years into a premier nuclear reaction data portal. This system is based on the 
latest computer technologies and  is well-positioned  to satisfy  current  and future nuclear science and technology needs and challenges.

\section{Acknowledgments}
\label{sec:Acknowledgements}
EXFOR is a large nuclear physics project that includes contributions from many responsible parties over the last 65 years, and we are grateful to the NRDC network members for the tireless work on EXFOR data compilations and overall database improvements. We would like to acknowledge V.P. Pronyaev, O. Schwerer (IAEA) and the late V. McLane (BNL), for productive discussions and constructive contributions, and V. Unferth (Viterbo University) for a careful reading of the manuscript and valuable suggestions. We appreciate useful consultations with K.I. Zolotarev (IPPE, Obninsk) and S. Simakov (Forschungszentrum Karlsruhe) on data renormalization and inverse reaction calculations, respectively. This work at Brookhaven National Laboratory was sponsored in part by the Office of Nuclear Physics, Office of Science of the U.S. Department of Energy under Contract No. DE-AC02-98CH10886 with Brookhaven Science Associates, LLC.







\begin{thebibliography}{99}


\bibitem{Holden05} N. Holden,   ``A Short History of CSISRS: At the Cutting Edge of Nuclear Data Information Storage and Retrieval Systems and its Relationship 
to CINDA, EXFOR and ENDF (2005)." Available from $\langle$http://www.nndc.bnl.gov/exfor/compilations/CSISRSHistory.pdf$\rangle$. 

\bibitem{Otuka14} N. Otuka, E. Dupont, V. Semkova  et al.,``Towards a More Complete and Accurate Experimental Nuclear Reaction Data Library (EXFOR): 
International Collaboration between Nuclear Reaction Data Centres (NRDC)." {\sc Nucl. Data Sheets} {\bf 120} 272, (2014).


\bibitem{Pritychenko15} B. Pritychenko, ``Intriguing Trends in Nuclear Physics Authorship." {\sc Scientometrics} {\bf 105} 1781, (2015).

\bibitem{Pearl70} S. Pearlstein, ``The NNCSC: Its History and Functions," {\sc Nucl. News}, p. 73 - 76 (1970). 

\bibitem{SG30}  A. Koning, A. Mengoni, E. Dupont et al., `` Quality Improvement of the EXFOR database,Ó Nuclear Energy Agency NEA/WPEC-30, NEA/NSC/WPEC/DOC 428 (2010).


\bibitem{Kam72} J.L. Kammerdiener, ``Neutron Spectra Emitted by $^{239}$Pu, $^{238}$U, $^{235}$U, Pb, Nb, Ni, Al and C Irradiated by 14 MeV Neutrons,"  Ph.D. Thesis, University of California,  Davis (1972). Lawrence Livermore National Laboratory Report UCRL-51232 (1972).

\bibitem{Pik16} G. Pikulina, S. Taova, ``Development of Compilation Tools for the EXFOR Library," NRDC Working Paper WP2016-37 (2016). 

\bibitem{Mak13} A. Makinaga, ``New Functions of GSYS and Compilation Tools," EXFOR Compilation Workshop, IAEA, Vienna, August  27-30 (2013). Available from $\langle$http://www.jcprg.org/gsys/$\rangle$.


\bibitem{GND} C.M. Mattoon, B.R. Beck, N.R. Patel et al., ``Generalized Nuclear Data: A New Structure (with Supporting Infrastructure) for Handling Nuclear Data," {\sc Nucl. Data Sheets} {\bf 113},  3145 (2012).

\bibitem{VMclane}   V. McLane, ``EXFOR Basics, A Short Guide to the Nuclear Reaction Data Exchange Format," Brookhaven National Laboratory Report BNL-NCS-63380-2000/05-Rev. (2000). 
O. Schwerer, ``EXFOR Formats Description for Users (EXFOR Basics)", International Atomic Energy Agency Report IAEA-NDS-206, June 2008. 
Available from  $\langle$https://www-nds.iaea.org/nrdc/nrdc\_doc/iaea-nds-0206-200806.pdf$\rangle$.

\bibitem{Otto15}  O. Schwerer, ``LEXFOR (EXFOR Compiler's Manual)", International Atomic Energy Agency Report IAEA-NDS-208, Rev. 2015/08 (2015). 
Available from $\langle$https://www-nds.iaea.org/nrdc/nrdc\_doc/iaea-nds-0208-201508.pdf$\rangle$.

\bibitem{Schw15}  O. Schwerer, ``EXFOR Formats Manual," International Atomic Energy Agency Report  IAEA-NDS-207, Rev. 2015/08 (2015).  Available from  $\langle$https://www-nds.iaea.org/nrdc/nrdc\_doc/iaea-nds-0207-201508.pdf$\rangle$.

\bibitem{Cullen86}  D.E.Cullen, A.Trkov, ``PROGRAM X4TOC4 (Version 2001-3) Translation of Experimental Data from the EXFOR Format to a Computation Format," 
International Atomic Energy Agency Report NDS-80, Rev.1 March 2001.  Available from  $\langle$https://www-nds.iaea.org/publications/iaea-nds/iaea-nds-0080.pdf$\rangle$.

\bibitem{Sch14}  O. Schwerer, N. Otuka, ``EXFOR/CINDA Dictionary Manual," International Atomic Energy Agency Report  IAEA-NDS-213, Rev. 2014/12 (2014).   Available from  $\langle$https://www-nds.iaea.org/nrdc/nrdc\_doc/iaea-nds-0213-201412.pdf$\rangle$.

\bibitem{Tech:01} V. Pronyaev, D. Winchell, V. Zerkin  {\it et al.}, ``Requirements for the Next Generation of Nuclear Databases and Services," {\sc J. Nucl. Sci. Tech.} Suppl. 2, 1476 (2002).   


\bibitem{NNDCWork:00} Proc. of Workshop on Relational Database and Java Tech. for  Nuclear  Data, BNL, September 11-15, 2000,  edited by R. Arcilla, D. Winchell, Y. Sanborn, unpublished. 

\bibitem{Zerkin01} V. Zerkin, ``EXFOR as a Multi-Platform Relational Database: Current Status of Development," NRDC Working Paper WP2001-25 (2001).   

\bibitem{pr06} B. Pritychenko, A.A. Sonzogni, D.F. Winchell {\it et al.}, ``Nuclear Reaction and Structure Data Services of the National Nuclear Data Center," {\sc Ann. Nucl. Energy} {\bf 33}, 390 (2006).

\bibitem{pr11} B. Pritychenko, E. Betak, M.A. Kellett, B. Singh, J. Totans, `` The Nuclear Science References (NSR) Database and Web Retrieval System," {\sc Nucl. Instr. and Meth.} {\bf A 640}, 213 (2011).

\bibitem{pr08} B. Pritychenko, A.A. Sonzogni, ``Sigma: Web Retrieval Interface for Nuclear Reaction Data," {\sc Nucl. Data Sheets} {\bf 109}, (2008) 2822.     

\bibitem{Zerkin05}  V.V. Zerkin,  V. McLane, M.W. Herman, C.L. Dunford,  ``EXFOR-CINDA-ENDF: Migration of Databases to Give Higher-Quality Nuclear Data Services." 
{\sc AIP Conf. Proc. of Intern. Conf. on Nucl. Data for Science and Technology} {\bf 769} 586, (2005); Santa Fe, New Mexico 2004.

\bibitem{ZVView} V. Zerkin, ``Interactive  Visual  Analysis  of  Remote/Local  Nuclear  Data  with  ZVView,"  pp.  45-48  in  Summary  
Report  on  Technical  Meeting  on  ``International  Reactor  Dosimetry  File:  IRDF-2002,"  Vienna,  Austria,  27-29  August 2002, INDC(NDS)-435 (2002); 
``DINAMO/ZVView Software for Interactive Visual Analysis of Nuclear Data," NRDC Working Paper WP1999-18 (1999).  Available from $\langle$https://www-nds.iaea.org/public/zvview/$\rangle$.

\bibitem{endf} A. Trkov, M. Herman, D.A. Brown, ``ENDF-6 Formats Manual," Brookhaven National Laboratory Report BNL-90365-2009 (2011). 

\bibitem{Chadwick11}  M.B. Chadwick, M. Herman, P. Oblo\v{z}insk\'{y}, et al.,  ``ENDF/B-VII.1 Nuclear Data for Science and Technology: Cross Sections, Covariances, Fission Product Yields and Decay Data,"  {\sc Nucl. Data Sheets} {\bf 112},  2887 (2011).

\bibitem{11Kon} A. J. Koning, E. Bauge, C.J. Dean {\it et al.}, ``Status of the JEFF Nuclear Data Library," {\sc J. Korean Physical Society}  {\bf 59}, No. 2, 1057 (2011).
\bibitem{15Kon}  A.J. Koning, D. Rochman, J. Kopecky {\it et al.},  ``TENDL-2015: TALYS-based Evaluated Nuclear Data Library," Available from $\langle$https://tendl.web.psi.ch/tendl\_2015/tendl2015.html$\rangle$. 
\bibitem{11Shi} K. Shibata, T. Kawano, T. Nakagawa {\it et al.}, ``JENDL-4.0: A New Library for Nuclear Science and Engineering," {\sc J. Nucl. Science and Technology} {\bf 48}, 1 (2011). 
\bibitem{07Zab} S.V. Zabrodskaya, A.V. Ignatyuk , V.N. Koscheev {\it et al.}, ``ROSFOND - Rossiyskaya Natsionalnaya Biblioteka Nejtronnykh Dannykh," {\sc VANT}, Nuclear Constants {\bf 1-2}, 3 (2007).
\bibitem{11Ge} Z.G. Ge, Z.X. Zhao, H.H. Xia {\it et al.}, ``The Updated Version of Chinese Evaluated Nuclear Data Library (CENDL-3.1)," J. Korean Physical Society 59 (2), 1052 (2011).



\bibitem{cinda} Computer Index of Nuclear Reaction Data (CINDA), Available from $\langle$http://www-nds.iaea.org/cinda$\rangle$.


\bibitem{cindam} H. Henriksson, ``CINDA Compilers Manual," NEA Databank document NEA/DB/DOC(2008)3, May 14 (2008).

\bibitem{NEA}  N. Soppera, M. Bossant, O. Cabellos, E. Dupont, C.J. Diez, ``JANIS: NEA JAva-based Nuclear Data Information System," {\sc EPJ Web Conf.} {\bf 146},   07006 (2017).

\bibitem{NRDC} Network of Nuclear Reaction Data Centres (NRDC) network, Available from $\langle$https://www-nds.iaea.org/nrdc/$\rangle$.


\bibitem{Zer07} V. Zerkin, A. Trkov, ``Development of IAEA Nuclear Reaction Databases and Services," {\sc Proc. Inter. Conf. on Nucl. Data for Sci. and Technology} April 22-27, 2007, Nice, France, editors O. Bersillon, F. Gunsing, E. Bauge, R. Jacqmin, and S. Leray; EDP Sciences, 769, (2008).

\bibitem{Has08} A. Hasegawa, H. Henriksson, F.J. Mompean {\it et al.}, ``Nuclear data activities at the NEA Data Bank," {\sc Proc. Inter. Conf. on Nucl. Data for Sci. and Technology} April 22-27, 2007, Nice, France, editors O. Bersillon, F. Gunsing, E. Bauge, R. Jacqmin, and S. Leray; EDP Sciences, 206, (2008).

\bibitem{Mom16} T.J. Mertzimekis, K. Stamou, A. Psaltis, ``An Online Database of Nuclear Electromagnetic Moments," {\sc Nucl. Instrum. Meth. Phys. Res.} {\bf A 807}, 56 (2016).

\bibitem{NRV17}  A.V. Karpov, A.S. Denikin, M.A. Naumenko {\it et al.},  ``NRV Web Knowledge Base on Low-Energy Nuclear Physics," {\sc Nucl. Instrum. and Meth. Phys. Res.} {\bf  A 859}, 112 (2017).

\bibitem{apache} The Apache Software Foundation, Available from $\langle$http://www.apache.org$\rangle$.

\bibitem{tomcat} The Apache Jakarta project, Available from $\langle$http://www.jakarta.apache.org$\rangle$.

\bibitem{ibandl}  Ion Beam Analysis Nuclear Data Library (IBANDL), Available from $\langle$https://www-nds.iaea.org/ibandl$\rangle$.

\bibitem{pr12} B. Pritychenko, M. Herman, ``National Nuclear Data Center: A Worldwide User Facility," {\sc Nucl. Phys. News} {\bf 22}, Issue 3, 23 (2012).


\bibitem{2007EMPIRE} M. Herman, R. Capote, B.V. Carlson {\it et al.}, ``EMPIRE: Nuclear Reaction Model Code System for Data Evaluation,"  {\sc Nucl. Data Sheets} {\bf 108}, 2655 (2007).

\bibitem{2012TALYS}  A.J. Koning, D. Rochman, ``Modern Nuclear Data Evaluation with the TALYS Code System," {\sc Nucl. Data Sheets} {\bf} 113, 2841 (2012).

\bibitem{GEANT}  J. Allison, K. Amako, J. Apostolakis {\it et al.}, ``Recent Developments in Geant4," {\sc Nucl. Instrum. Meth. Phys. Res.} {\bf A 835}, 186 (2016).

\bibitem{MCNP}  X-5 Monte Carlo Team, ``MCNP - A General Monte Carlo N-Particle Transport Code, Version 5," Los Alamos National Laboratory Report LA-UR-03-1987 (2003). Available from $\langle$https://laws.lanl.gov/vhosts/mcnp.lanl.gov/pdf\_files/la-ur-03-1987.pdf$\rangle$.

\bibitem{Prepro}  D.E. Cullen, `` PREPRO 2015, 2015 ENDF/B Pre-processing Codes,"  International Atomic Energy Agency Report IAEA-NDS-39 (2015); Rev. 16, January 31, 2015. Available from $\langle$https://www-nds.iaea.org/public/endf/prepro/$\rangle$.

\bibitem{ensdf} T.W. Burrows, ``The evaluated nuclear structure data file: Philosophy, content, and uses," {\sc Nucl. Instrum. and Meth. Phys. Res.} {\bf  A 286}, 595 (1990). Available from $\langle$http://www.nndc.bnl.gov/ensdf$\rangle$.

\bibitem{Zerkin12} V.V. Zerkin, ``EXFOR Data Correction System: Progress in 2011-2012," NRDC Meeting, Paris (2012).

\bibitem{ZolCor}  K.I. Zolotarev, ``New Russian Evaluations for IRDF-2002," International Reactor Dosimetry File 2002 (IRDF-2002), IAEA Technical Reports Series No. 452 (2006).

\bibitem{Zup80} E. Zupranska, K. Rusek, J. Turkiewicz, P. Zupranski, ``Excitation Functions for (n,$\alpha$) Reactions in the Neutron Energy Range from 13 to 18 MeV," {\sc Acta Phys. Pol.} {\bf B11}, 853 (1980).

\bibitem{Carl09} A.D. Carlson, V.G. Pronyaev, D.L. Smith {\it et al.}, ``International Evaluation of Neutron Cross Section Standards,"  {\sc Nucl. Data Sheets} {\bf 110}, 3215 (2009).

\bibitem{Zerkin121} V. Zerkin, ``Web Tool for Constructing a Covariance Matrix from EXFOR Uncertainties," {\sc EPJ Web of Conferences} {\bf 27}, 00009 (2012).

\bibitem{Sage10} C. Sage, V. Semkova, O. Bouland {\it et al.}, ``High resolution measurements of the 241Am(n,2n) reaction cross section," 
{\sc Phys. Rev.} {\bf C 81}, 064604 (2010).

\bibitem{88Rol} C.E. Rolfs and W.S. Rodney, {\sc Cauldrons in the Cosmos}, The University of Chicago Press (1988).

\bibitem{08Lar} N.M. Larson, ``Updated Users Guide for SAMMY: Multilevel R-Matrix Fits to Neutron Data Using Bayes Equations," Oak Ridge National Laboratory Report ORNL/TM-9179/R8 (2008).

\bibitem{Har05}  S. Harissopulos, H.W. Becker, J.W. Hammer et al., ``Cross Section of the $^{13}C(\alpha,n)^{16}O$ Reaction: A Background for the Measurement of Geo-neutrinos," {\sc Phys. Rev.} {\bf C 72}, 062801 (2005).

\end{thebibliography}



\end{document}